*To be submitted to Ceramics International*

# Pyroelectric thermo-wave probing of $BaZr_{0.2}Ti_{0.8}O_3$ ceramics

Nicholas V. Morozovsky[1], Yuriy O. Zagorodniy[2], Mykola Yelisieiev[3,1], Oleksandr S. Pylypchuk[1], Mykola O. Semenenko[3,1], Iryna V. Kondakova[2], Lesya P. Yurchenko[2], Viktor Bovtun[4], Valentin V. Laguta[2,4], Eugene A. Eliseev[2*], Volodymyr M. Dzhagan[3†], and Anna N. Morozovska[1‡]

[1] *Institute of Physics, National Academy of Sciences of Ukraine, 46, Nauki Avenue, 03028 Kyiv, Ukraine*

[2] *Frantsevich Institute for Problems in Materials Science, National Academy of Sciences of Ukraine, 3, str. Omeliana Pritsaka, 03142 Kyiv, Ukraine*

[3] *V. Lashkaryov Institute of Semiconductor Physics, National Academy of Sciences of Ukraine, 41, Nauki Avenue, 03028 Kyiv, Ukraine*

[4] *Institute of Physics, Czech Academy of Sciences, Na Slovance 1999/2, 18200 Prague, Czech Republic*

## Abstract

$Ba_xZr_{1-x}TiO_3$ ceramics with $0.15 < x < 0.25$, which have high dielectric permittivity, small leakage currents and low dissipation factor, are promising materials for tunable capacitor devices, multilayered ceramic capacitors and piezoelectric actuators. These materials reveal relaxor properties due to the local chemical strains caused by the isovalent substitution of $Ti^{4+}$ ions by $Zr^{4+}$ ions with a larger ionic size. However, to the best of our knowledge, the pyroelectric properties of $Ba_xZr_{1-x}TiO_3$ ceramics with x~0.2 have not been studied. To fill the gap in knowledge, we perform the pyroelectric thermo-wave probing of $BaZr_{0.2}Ti_{0.8}O_3$ ceramics prepared by the solid-state synthesis. Results of the thermo-wave probing of the most interesting polar states, namely polarized, depolarized, relaxed and restored polarized states, reveal the pronounced pyroelectric response that can be strongly asymmetric with respect to the opposite surfaces of the ceramics. The asymmetry and profiles of pyroelectric response depend significantly on the pre-history of the electric field cycling indicating possible non-ergodic relaxor-type polar states in the ceramic sample. X-ray diffraction spectrum, recorded at room temperature, reveals the virtual absence of the macroscopic tetragonality and the weak asymmetry of the (200) peak, which indicate the small tetragonality inside the Ti-enriched polar nanoregions. Decrease of the relative intensity of the Raman band at 714 $cm^{-1}$ occurring upon heating evidences the diffuse ferroelectric-paraelectric phase transition between 30°C and 40°C. The temperature dependence of the dielectric permittivity obeys modified Curie-Weiss law with power 1.4, indicating that both ordered ferroelectric and relaxor states coexist in the $BaTi_{0.8}Zr_{0.2}O_3$ ceramics. Estimated pyroelectric coefficients, which are ~$5\cdot10^{-4}$ $C/m^2K$ in the ordered ferroelectric state and at least 10 times larger in the relaxor ferroelectric state, are relatively high. Therefore, obtained results can be useful for elaboration of lead-free relaxor ferroelectric ceramics for advanced pyroelectric and electrocaloric applications.

[*] corresponding author, e-mail: eugene.a.eliseev@gmail.com

[†] corresponding author, e-mail: volodymyrdzhagan@gmail.com

[‡] corresponding author, e-mail: anna.n.morozovska@gmail.com

## 1. INTRODUCTION

Ferroelectric compounds of the perovskite structure with chemical formulae $ABO_3$, where A and B are cations, are of permanent fundamental and applied interest. A great interest is devoted to the lead-free $ABO_3$ compounds, such as single-crystals, nanoparticles and/or fine-grained ceramics based on barium titanate $BaTiO_3$ (BT), which are well-known indispensable classical materials for piezoelectric transducers [1, 2, 3], pyroelectric detectors [4, 5, 6, 7], electrocaloric converters [8, 9, 10], energy storage and harvesting [11, 12, 13], as well as multilayered ceramic capacitors (MLCCs) [14, 15]. As a rule, $BaTiO_3$-based systems are dielectrics with a high permittivity [16, 17] and tunability [18], whose temperature of the ferroelectric phase transition and dielectric properties can be tuned by isovalent substitution of A ($Ba^{2+}$) and/or B ($Ti^{4+}$) sites by Ca, Sn, Sr, Zr and/or Eu (see, e.g., Chapter 5 in Ref. [1]).

During the past decades, many studies were focused on preparation, microstructure, dielectric, and ferroelectric properties of A-A-substituted $Ba_xSr_{1-x}TiO_3$ (BST) and B-B-substituted $Ba_xZr_{1-x}TiO_3$ (BZT) single-crystals, micro- and nanoparticles, and dense fine-grained ceramics. In the case of BST, $Ba^{2+}$ ions (ionic radius 0.135 nm) are replaced by smaller $Sr^{2+}$ ions (ionic radius 0.118 nm). In the case of BZT, $Ti^{3+}$ and $Ti^{4+}$ ions (with radii of 0.067 nm 0.0605 nm, respectively) are replaced by larger $Zr^{4+}$ ions (ionic radius 0.072 nm, according to the data in Ref. [19]). Partial A-A-substitution of $Ba^{2+}$ by $Sr^{2+}$ in the case of BST and the B-B-substitution of $Ti^{4+}$ by $Zr^{4+}$ in the case of BZT leads to the deformations of $TiO_6$-octahedra, and so, to the local strains in the BTO lattice [18, 20]. This leads to decrease of the lattice constant *a* for BST [21], and to increase of lattice constants *a* and *c* for BZT [22]. The changes of lattice constants are followed by the controllable decrease in the ferroelectric-to-paraelectric transition temperature for both BST [21] and BZT [16, 22] compounds. Due to this, the A-A-substituted BST compounds, which have high dielectric permittivity, small leakage currents and low dissipation factor [23, 24, 25], are considered as advanced materials for tunable capacitor devices [26, 27] and MLCC [28] applications. Along with this, the B-B-substituted BZT compounds have been shown as an alternative to the BTO and BST for applications in MLCCs [29, 30, 31], tunable capacitor devices [18, 32, 33] and piezoelectric actuators [34, 35], because of the suppressed electron hopping between $Ti^{4+}$ and $Ti^{3+}$ ions due to the partial isovalent substitution of $Ti^{4+}$ ions by $Zr^{4+}$ ions.

Depending on the x value, the $BaTi_{1-x}Zr_xO_3$ compounds, either single crystals [34, 36] or ceramics [16, 37], undergo the ferroelectric-paraelectric phase transition of the first or the second order at certain temperature. At room temperatures they belong to a rhombohedral, orthorhombic or tetragonal structure depending on the x-value. With the increase in the x value, the three phase transition points between the rhombohedral, orthorhombic, tetragonal and cubic phases of BZT with come closer to each other and merge together for $0.13 < x < 0.15$ [16, 20, 22, 34 - 38, 39]. With the increase of the Zr content x the transition temperature between the ferroelectric tetragonal and the paraelectric cubic phases decreases in BZT compounds [16, 40]. At the same time, the diffuseness of the phase transition increases with increase in x, so that the temperature behaviour of the dielectric permittivity of BZT ceramics changes from characteristic for "ordered" ferroelectrics at $x < 0.15$ to the one characteristic for "disordered" relaxor-type ferroelectrics at $x > 0.15$ [16, 18, 20, 35, 40, 39, 41].

The $BaZr_{0.2}Ti_{0.8}O_3$ (BZT-20) compound with x = 0.2 possesses an appreciably diffuse ferroelectric-paraelectric phase transition [16, 18], which is close to a boundary between the first and the second order phase transitions, $x \approx 0.15$ [16, 39]. The x-value is close to a Zr-concentration threshold between ordered ferroelectric and relaxor-like ferroelectric behaviour [18, 20, 33, 39]. For BZT-20, a diffuse maximum of the dielectric permittivity $\varepsilon(T)$ is observed in the vicinity of $T_m =$ 300 K [16, 18, 22, 37]. As it was shown in Ref. [16], there exists a diffused transition with an inflection point on the temperature dependence of the remanent polarization $P_r(T)$ in the vicinity of 20ºC for BZT-20 (see Fig. 5 in Ref. [16]). The behavior of $\varepsilon(T)$ of BZT with $x > 0.15$ in the vide vicinity of its broad temperature maximum around $T_m$ is characteristic for relaxor ferroelectrics [18, 40]. For RFEs the localization of polarization within disordered dynamic polar nanoregions (PNRs) irregularly embedded in a non-polar (paraelectric) surrounding is inherent [20, 30, 33]. The reason of the relaxor-like behavior of BZT compounds with $x > 0.15$ is related to the existence of polar nanosized clusters enriched in Ti (see **Scheme 1(a)** and Refs. [18, 33]), and non-polar or weakly polar clusters enriched in Zr (see **Scheme 1(b)** and Refs. [18, 33, 42]). In BZT, due to the irregular spatial distribution of those clusters, which differ in size and dipole moment magnitude and direction [18, 33], each polar cluster can be assigned to an individual PNR as in relaxor ferroelectrics [20, 30, 33, 42].

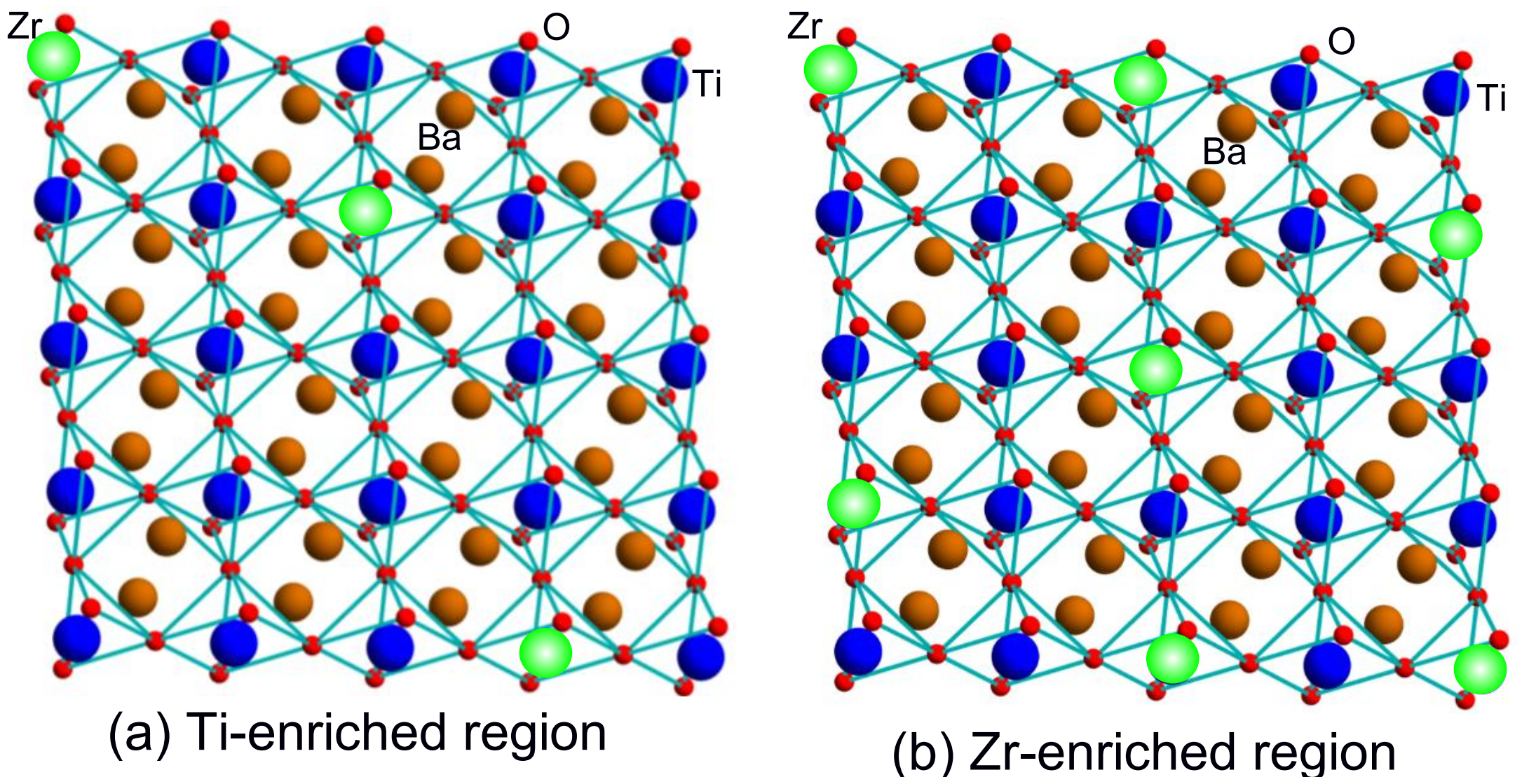

**SCHEME 1.** Two fragments of the $BaZr_{0.2}Ti_{0.8}O_3$ lattice corresponding to the Ti-enriched **(a)** and Zr-enriched **(b)** regions. The stoichiometric ratio of Ti:Zr ions is 5:1.

Based on the results of X-ray diffraction and Raman spectroscopy it was concluded [42] that the appearance and growth of Ti-enriched PNRs in $BaZr_xTi_{1-x}O_3$ compounds with increase in $x$ is related with an interplay between $BaTiO_3$ ferroelectric and $BaZrO_3$ paraelectric sublattices leading to the segregation of Zr ions outside the Ti-enriched clusters with the reduction of the local strains, arising due to the difference in $Ti^{4+}$ and $Zr^{4+}$ ionic radii ($r_{Zr}^{4+}/r_{Ti}^{4+}$ = 1.19 [39]). As a result, a heterogeneous set of the Ti-polar core – Zr-non-polar shell type nanostructures arises, where the segregation of $Zr^{4+}$ ions is a factor limiting the growth of Ti-enriched PNRs [20, 42]. In this case, the ferroelectrically active $Ti^{4+}$ ions are not centred in the $TiO_6$ octahedra, whereas the $Zr^{4+}$ ions are centred in the $ZrO_6$ octahedra in BZT [20].

Following Refs [20, 43], it can be assumed that the $Ti^{4+}$ ions, located inside polar Ti-enriched clusters, can randomly occupy one of the eight equivalent off-centre positions along the [111] cubic directions in $BO_6$ octahedra of BZT, similarly to the elementary cell of BTO. Therefore, BZT ceramics, where polar Ti-enriched clusters are differently strained due to different $Zr^{4+}$ surroundings [42], can possess different local piezoelectric and pyroelectric response. Considering the reasons, lead-free $BaZr_{0.03}Ti_{0.97}O_3$ ceramics is a promising material for mechanical to electrical energy conversion; they have good piezoelectric performances at low sound frequencies comparable with those of lead-based $PbZr_xTi_{1-x}O_3$ ceramics [44]. Therefore, it can be expected that the pyroelectric performances of BZT ceramics, which is determined mainly by the polar reaction of dipoles in Ti-enriched clusters to temperature changes, will be similar to the pyroelectric response of thermally excited PNR dipoles in relaxors, in particular in the lead magno-niobate, $Pb(Mg_{1/3}Nb_{2/3})O_3$.

To the best of our knowledge, pyroelectric performances were studied only for [(1-x)$Ba_{0.9}Ca_{0.1}TiO_3$-x($BaSn_{0.2}Ti_{0.8}O_3$)] [45] and ($Ba_{0.825+x}Ca_{0.175-x}$)($Ti_{1-x}Sn_x$)$O_3$ [46] ceramics with x = 0.6 and x = 0.0625, respectively. It was shown that both ceramics could be of potential interest for pyroelectric applications [45, 46], at that the maximal "effective" pyroelectric coefficient, $dP_r(T)/dT \cong 5.7 \cdot 10^{-4}$ C/m$^2$K at 332 K, was estimated from the temperature dependence of the remanent polarization $P_r(T)$ of ($Ba_{0.825+x}Ca_{0.175-x}$)($Ti_{1-x}Sn_x$)$O_3$ with x = 0.0625 [46]. However, to the best of our knowledge, the pyroelectric properties of BZT ceramics with x~0.2. have not been studied. To fill the gap in knowledge, we perform the pyroelectric thermo-wave probing of BZT-20 ceramics in this work.

## 2. CHARACTERIZATION OF CERAMIC SAMPLES AND METHODS

Relaxor-type BZT-20 dense ceramic samples were prepared by solid-state synthesis (see details in Ref. [47]). The ceramic samples were sintered as tablets with a thickness of 0.5 mm and a base area of 0.42 cm$^2$, polished on both sides. X-ray diffraction (XRD) measurements were carried out using a Philips X'Pert PRO – MRD diffractometer equipped with a CuKα radiation source (wavelength λ = 0.15406 nm). The anode voltage of the tube was 45 kV, current 40 mA. Diffraction patterns were registered in the symmetric (2θ/ω) geometry employing either standard or high-resolution mode. The obtained set of XRD reflexes, shown in **Fig. 1(a)**, agree with those previously reported for the fully depolarized BZT-20 ceramics in a (pseudo)cubic phase [21, 22, 27, 33]. Indeed, splitting of the (110), (200) or (210) reflections are virtually absent in **Fig. 1(a)**, indicating the absence of apparent effective tetragonality of the BZT-20 ceramics at 25$^0$C. Only the (200) peak is slightly asymmetric, which may be related to the presence of small tetragonality inside Ti-enriched PNRs. Inset to **Fig. 1(a)** shows a typical XRD spectra of $BaTiO_3$ nanopowders [48], where the splitting of (200) and (201) reflections is evident. As in the case of $BaTiO_3$ nanopowders, the high-accuracy synchrotron XRD spectroscopy can resolve very small splitting of higher-order reflections produced by the PNRs, but the technique is not available for the present work.

A valuable structural characterization, that evidences the diffuse ferroelectric-paraelectric phase transition near 300 K in the BZT-20 ceramics, has been provided by the Raman spectroscopy in the actual temperature range (see **Fig. 1(b)**). Raman spectra were excited with 532 nm solid-state laser and registered using a single-stage spectrometer MDR-23 (LOMO) equipped with a cooled CCD detector (Andor iDus 401, UK). The spectrum acquired at room temperature (upper or bottom green curves in **Fig. 1(b)**) coincides with those reported earlier for BZT-20 ceramics [42]. The decrease of the relative intensity of the Raman band at 714 cm$^{-1}$ with respect to other modes, which occurs upon heating (see inset to **Fig. 1(b)**), clearly demonstrates the ferroelectric-paraelectric phase transition occurring between 30°C and 40°C, in agreement with previous results [42].

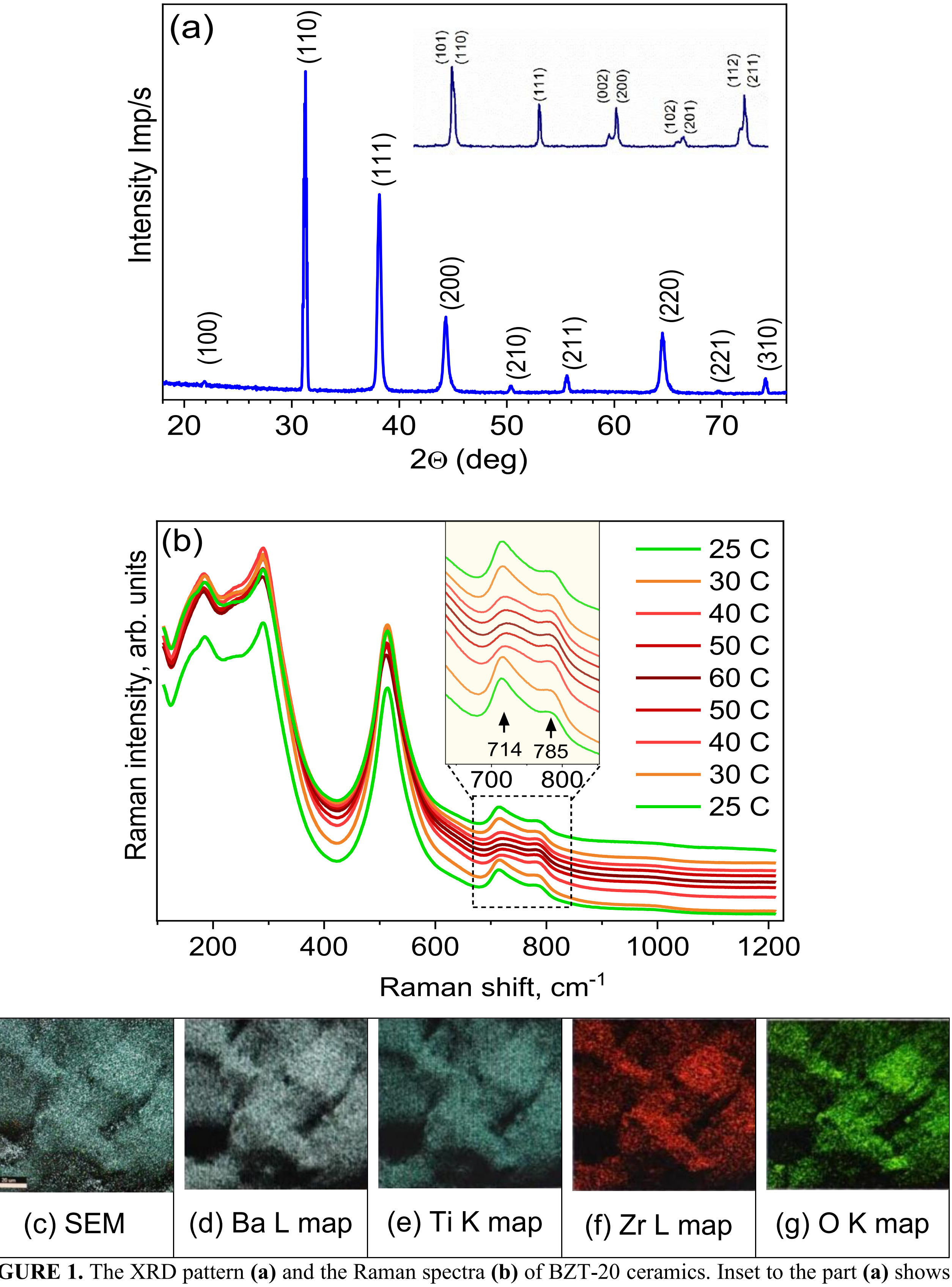

**FIGURE 1.** The XRD pattern **(a)** and the Raman spectra **(b)** of BZT-20 ceramics. Inset to the part **(a)** shows a typical XRD spectra of $BaTiO_3$ nanopowders adapted from Ref. [48]. **(c)** Typical SEM image of the BZT-20 ceramics**. (d)-(g)** Element maps, obtained by EDS analysis.

Notably, that the scanning electron microscopy (SEM) shows a typical grain-sensitive morphology (see **Fig. 1(c)**). The EDS analysis demonstrates the quasi-uniform distribution of Ba, Ti and Zr at micron scale (see **Figs. 1(d)-(f)**). The oxygen atomic percentage (47.6%), that map is shown in **Fig. 1(g)** is noticeably lower than expected for the ideal $ABO_3$-perovskite stoichiometry of 60%, which is typical for standardless EDS quantification at 20 keV due to matrix absorption effects, low X-ray yield for light elements, and potential surface contamination.

For electrophysical, dielectric and pyroelectric measurements, the tablet was polished and coated with a silver conductive layer, which was used as electrodes. To study the dielectric properties, the tablet was coated with a silver conductive layer and used as electrodes. The dielectric properties of BZT-20 ceramic were measured on a sample prepared in the form of a tablet with a thickness of 0.5 mm and a base area of 0.42 $cm^2$. The tablet was polished and coated with a silver conductive layer, used as electrodes. The frequency and temperature dependence of the real ε' and imaginary ε'' components of the dielectric permittivity was investigated by measuring the capacitance ($C$) and loss tangent ($\tan\delta$) of the sample. Measurements were performed using the E7-20 impedance analyzer. The dielectric permittivity was measured at frequencies 100 Hz – 200 kHz in the temperature range 150-450 K. The values of real $\varepsilon'$ and imaginary $\varepsilon''$ components of dielectric permittivity were calculated as $\varepsilon' = Cd/\varepsilon_0 S$, and $\varepsilon''/\varepsilon' = tan\delta$, where $d$ is the thickness and $S$ is the area of the tablet. The polarization-field hysteresis loops $P(E)$ were obtained on a home-made device by measuring the polarization currents arising from a voltage applied to the sample electrodes.

## 3. DIELECTRIC SPECTROSCOPY AND FERROELECTRIC HYSTERESIS

The temperature variation of the real ε' and imaginary ε'' components of the dielectric permittivity measured for the studied BZT-20 sample are presented in **Fig. 2**. The dielectric permittivity has a wide maximum near 305 K and a weak frequency dependence, which is typical for ferroelectrics with a diffuse phase transition [49].

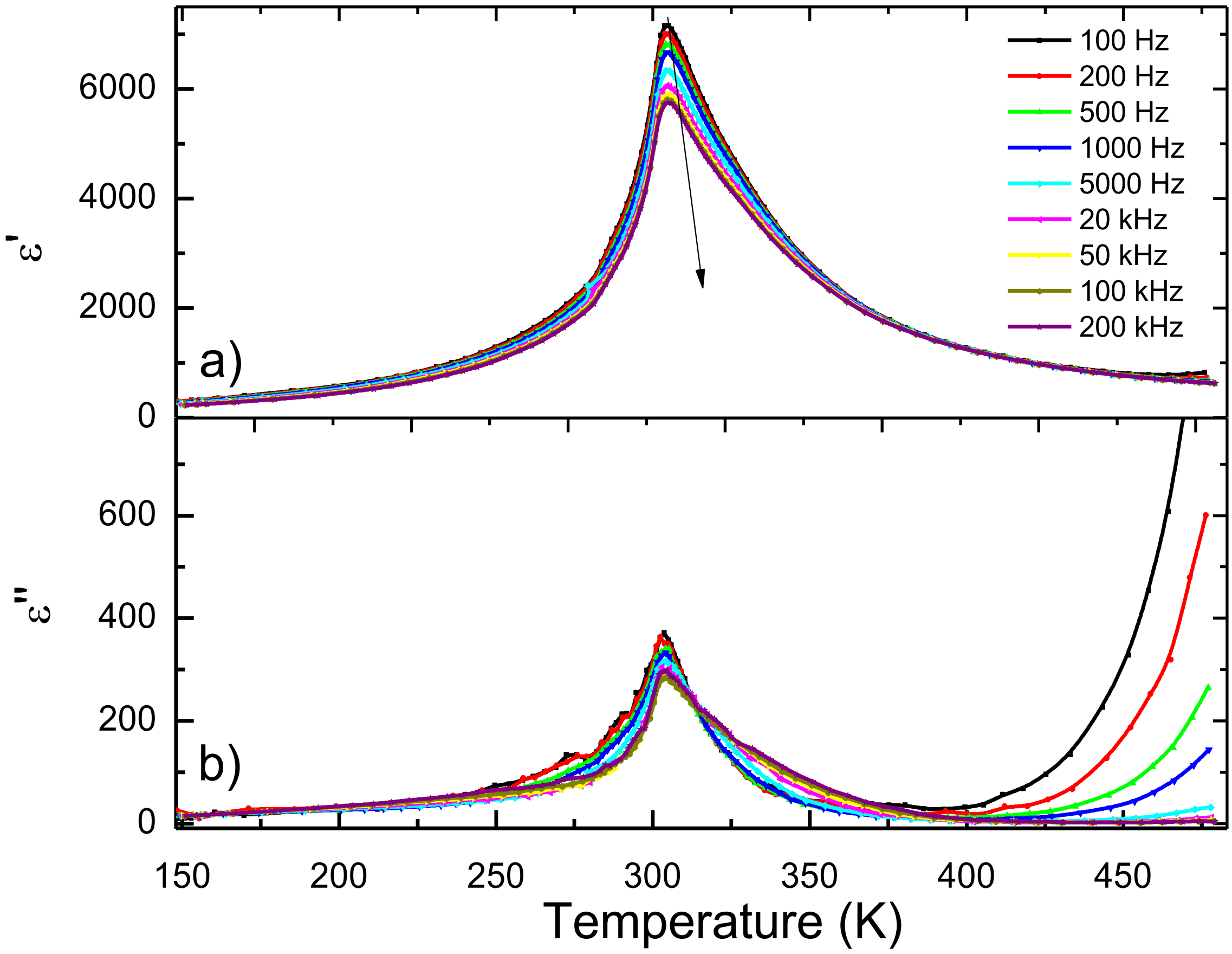

**FIGURE 2.** The temperature dependence of the real ε' **(a)** and imaginary ε" **(b)** parts of the relative dielectric permittivity of BZT-20 ceramics at different frequencies.

Since $Zr^{4+}$ ions have a larger ionic radius than $Ti^{4+}$ B-site lattice ions, isovalent substitution with Zr causes local chemical strains. Inhomogeneous chemical strains diffuse and broaden the ferroelectric-paraelectric transition temperature, making the system relaxor ferroelectrics. At the same time, Zr–O octahedra distort more strongly than Ti–O octahedra, creating additional local dipoles and possibly enhancing the dielectric permittivity at temperatures below the transition. At significant concentrations of Zr (~15 – 20 %), substituted ionic cites can increase the polaron conductivity at temperatures above the ferroelectric-paraelectric transition. Thus, an additional conductive mechanism arises, which explains the asymmetry of the dielectric permittivity and losses maxima shown in **Fig. 2**.

The temperature dependence of the dielectric permittivity of the compounds with a diffuse phase transition does not obey the Curie-Weiss law. For temperatures lying in a certain interval above the transition temperature $T_m$ it can be described using the modified Curie–Weiss law [50]:

$$\frac{1}{\varepsilon} = \frac{1}{\varepsilon_m} + \left(\frac{T - T_m}{C_{CW}}\right)^{\mu}, \quad (1)$$

Where $C_{CW}$ is the Curie–Weiss constant, $\varepsilon_m = \varepsilon'_{max}$, and the power $\mu$ characterizes the degree of phase transition diffuseness and can take values from 1 for ordered ferroelectrics to 2 for relaxors.

The temperature dependence of the inverse dielectric permittivity measured at 1 kHz is shown in **Fig. 3**. Fitting this data yields the value of $\mu$ approximately equal to 1.4. Such $\mu$-value indicates that the ordered ferroelectric state and the disordered relaxor state coexist in the ceramics [49], as well as suggests the presence of ferroelectric relaxor properties in the BZT-20 ceramics.

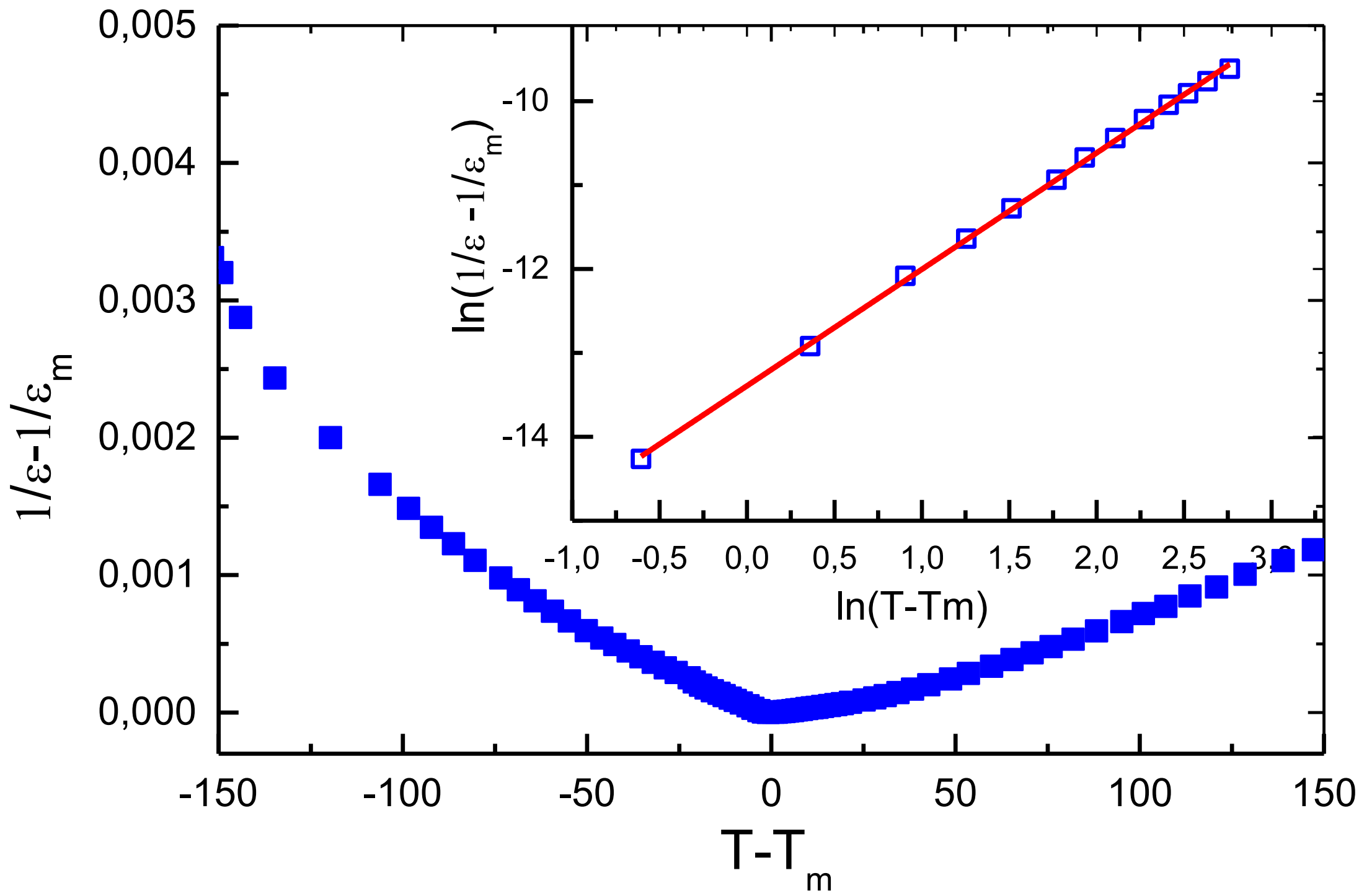

**FIGURE 3.** The temperature dependence of the inverse dielectric permittivity of BZT-20 ceramics measured at 1 kHz. Inset shows the logarithmic dependence of the difference $\frac{1}{\varepsilon} - \frac{1}{\varepsilon_m}$ on the reduced temperature $T - T_m$.

**Figure 4** shows the electric field dependence of the polarization $P(E)$ measured at room temperature. The presence of well-saturated hysteresis loops with the coercive field of 1.14 kV/cm and the remanent polarization of 6.7 μC/cm$^2$ confirms the presence of the ordered ferroelectric state in the pre-polarized BZT-20 ceramics.

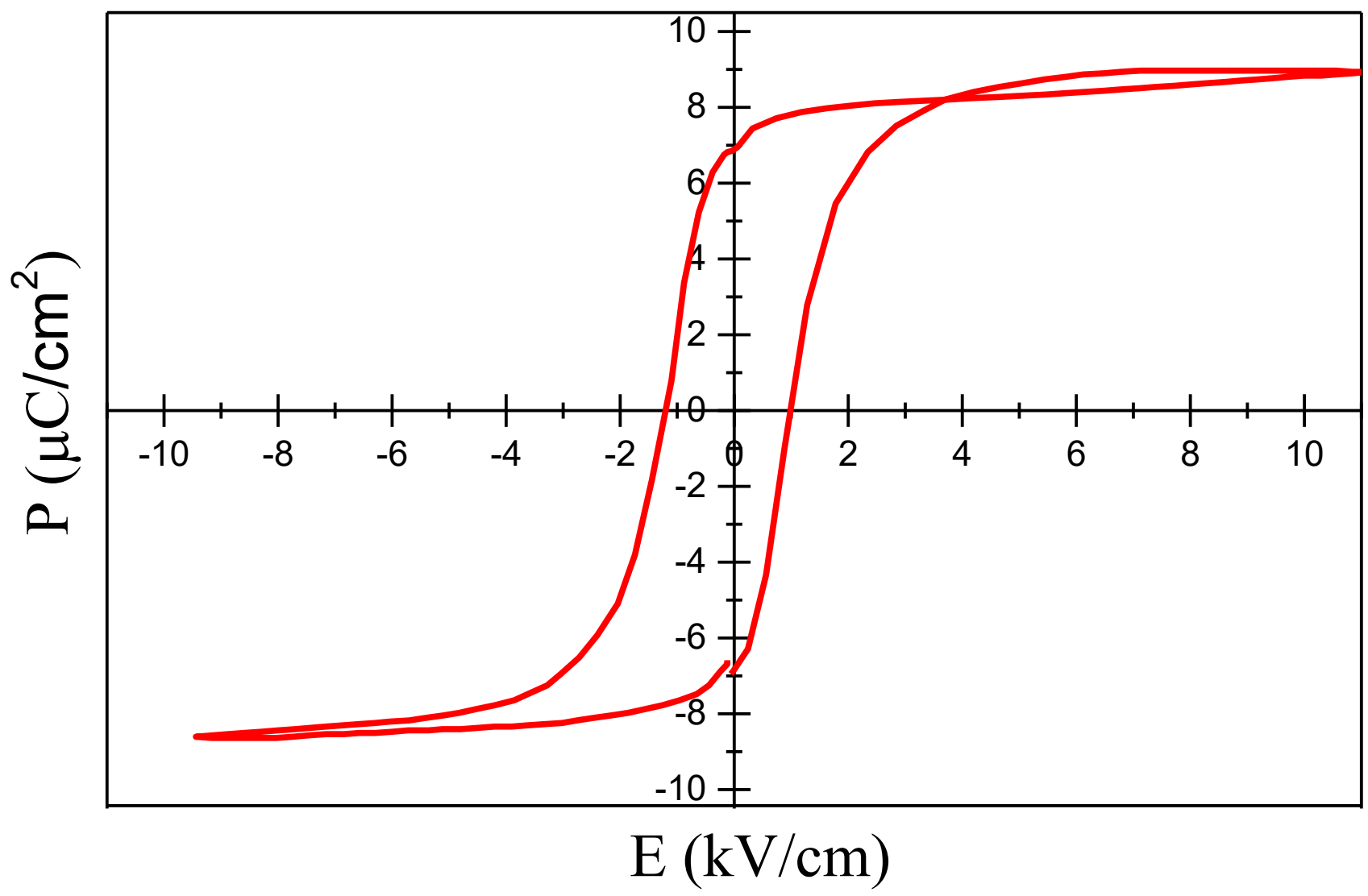

**FIGURE 4.** Typical $P(E)$ hysteresis loop of BZT-20 ceramics.

## 4. PYROELECTRIC THERMO-WAVE PROBING OF BZT-20 CERAMICS IN DIFFERENT POLARIZATION STATES

To clarify the polar state of the BZT-20, their pyroelectric response has been studied. The sample geometry corresponds to the Ag-paste electrodes of 0.5 mm thick, diameter D ≈ 8 mm disk (A ≈ 50 mm$^2$) with thin wires glued by the Ag-paste.

For pyroelectric studies, the photo-thermo-modulation pyroelectric method [51, 52] was used. In the method, the temperature of the sample changes as $T(t) = T_0 + \Delta T \sin(2\pi f_m t)$, where $\Delta T \ll T_0$ and $f_m$ is the modulation frequency of the thermal flux created by the IR-diode. The temperature change $\Delta T$ occurs under the influence of a sinusoidally modulated IR radiation flux with intensity $W(t) = \Delta W[1 + \delta \cdot \sin(2\pi f_m t)]$, where and the parameter δ can vary in the range $0 < \delta \leq 1$. To create the maximal modulation amplitude, the thermal flux of the IR-diode is sinusoidally modulated by the electric voltage with the frequency $f_m$. Since the flux is absent when the diode is closed, one can regard that $\delta \approx 1$ in the case considered. The measured value is the pyroelectric response $U_\pi$, caused by the temperature change of the spontaneous polarization $P_s(T)$ of the polar active material. The value of $U_\pi$ is proportional to the value of the pyroelectric coefficient $\gamma = dP_s(T)/dT$ [53].

The dynamic thermal excitation allows analyzing the amplitude-frequency, $U_\pi(f_m)$, and phase-frequency, $\varphi_\pi(f_m)$, dependences of $U_\pi$ in a wide frequency range. Under such conditions, it is possible to measure $U_\pi = U_{\pi 1}$ in the pyroelectric current mode, $U_\pi = U_{\pi 1}$, if $2\pi f_m R_L C_s \ll 1$ and in the pyroelectric voltage mode, $U_\pi = U_{\pi 2}$, if $2\pi f_m R_L C_s \gg 1$ [4]. Hereinafter, $R_L$ is the electrical resistance of the load in the circuit of the sensitive element, $C_s$ is the electrical capacitance of the element.

In the case of a uniform distribution of pyro-active regions over the thickness of the tablet, $U_{\pi 1,2}$ depends on the pyroelectric coefficient γ, volume heat capacity $c_T$, dielectric permittivity $\varepsilon_\pi$ of the ceramics and its thickness $d$ [4]. In the pyroelectric current mode, $U_{\pi 1} \propto \frac{\gamma}{c_T} \cdot \frac{R_L}{d}$ is constant and $\varphi_\pi = \varphi_{\pi 1}$ is constant, and those constants depend on the frequency $f_m$. Due to the resistive nature of the load in the sample circuit, $U_{\pi 1}(t)$ can be in-phase ($\varphi_{\pi 1} = 0$) or in anti-phase ($\varphi_{\pi 1} = 180^o$) with the IR flux intensity $W(t)$, depending on the direction of polarization in the sample. In the pyroelectric voltage mode, $U_{\pi 2} \propto \frac{\gamma}{c_T} \cdot \frac{1}{\varepsilon_\pi f_m}$, and $\varphi_\pi = \varphi_{\pi 2}$ is the $f_m$-dependent constant. As a result of the capacitive nature of the load in the sample circuit, there is a phase shift of ±π/2 (90° or 270°) between the $U_{\pi 2}(t)$ and the $W(t)$, depending on the polarization direction in the sample. It should be noted that the $U_{\pi 1}$ mode, chosen for the corresponding $f_m$ and $R_L$ so that $2\pi f_m R_L C_s < 1$, transforms into the $U_{\pi 2}$ mode when $2\pi f_m R_L C_s > 1$, and the transition frequency $f_{mt}$ is determined from the relation $2\pi f_{mt} R_L C_s = 1$.

From the explicit expression of the thermal diffusion length $\lambda_T = \sqrt{\frac{a_T}{\pi f_m}}$ ($a_T$ is the thermal diffusivity) it is possible to determine the features of the subsurface distribution of pyroelectric parameters, using the thermal wave scanning method [52]. Since any deviation from a uniform polarization distribution is reflected in the frequency dependences $U_{\pi 1,2}(f_m)$ and $\varphi_{\pi 1,2}(f_m)$, the corresponding $\lambda_T$-profiles of the pyroelectric response $U_{\pi 1,2}(\lambda_T)$ and the phase $\varphi_{\pi 1,2}(\lambda_T)$ can be obtained and analyzed [52].

The results of the thermos-wave probing of the most interesting (polarized, partially depolarized, semi-depolarized, relaxed depolarized, re-polarized and restored polarized) polarization states of BZT-20 ceramics measured at the room temperature ($T_0 \approx 25$°C) are presented in **Figs. 5-10**. The upper plots show the $U_\pi(f_m)$ dependences, and the middle plots show the dependences of the phase shift $\varphi_\pi(f_m)$ between $U_\pi(t)$ and $W(t)$, with the bottom panels showing the dependences of the so-called "dielectric ratio", $D_\pi = \frac{U_{\pi 1}}{U_{\pi 2} f_m} \sim \varepsilon_\pi$. The left and right columns of **Figs. 5 – 10** show the results of probing from the left surface and from the right surface of the BZT-20 pellet, respectively. For all polarization states, shown in **Figs. 5 – 10**, the pyroelectric response phase shift $\varphi_{\pi 1,2}(f_m)$ changes by more (or less) than 180 degrees at opposite surfaces of the pellet, which confirms the existence of the spatially inhomogeneous ferroelectric-like state of the ceramics.

Pyroelectric response of the *initial polarized state* of the BZT-20 ceramic pellet is shown in **Fig. 5**. Probing opposite surfaces of the sample, near equal $U_{\pi 1,2}$ values, as well as similar $U_{\pi 1,2}(f_m)$ and $\varphi_{\pi 1,2}(f_m)$ dependences, were obtained. General decrease of $U_{\pi 1} \sim \gamma$ and $f_m U_{\pi 2} \sim \gamma / \varepsilon_\pi$ values with the increase in $f_m$ indicates the gradual decrease of γ and $\gamma/\varepsilon_\pi$ values from the volume to the surface.

The peculiarities of $f_m U_{\pi 2}(f_m)$ , $\varphi_{\pi 1,2}(f_m)$ and $D_\pi(f_m) \sim \varepsilon_\pi$ in the vicinity of $f_m \approx 200$ Hz are most likely associated with the small subsurface inhomogeneities of the ratio $\gamma/\varepsilon_\pi$ located under the opposite surfaces of the sample at the thermal wave depth $\lambda_T \approx 20$ μm. The pyroelectric coefficient $\gamma$, estimated in the polarized state, is about $(0.5 - 5) \cdot 10^{-4}$ C/m$^2$K depending on the frequency $f_m$, that is comparable with the $\gamma$ values characteristic for BST ceramics [45, 46].

Then, we reversed the polarization in the sample and measured its pyroelectric response in the so-called "*partially*" *depolarized state*. This state is weakly asymmetric with respect to the opposite surfaces state was achieved after the application of 10 Hz sine voltage with amplitude ±300 V and its subsequent gradual vanishing to zero voltage. Pyroelectric response of the partially depolarized state of the BZT-20 ceramic pellet is shown in **Fig. 6**. The $U_{\pi 1,2}$ and $f_m U_{\pi 2}$ values, and the values of $\gamma$ and $\gamma/\varepsilon_\pi$, are approximately 2 times smaller than for the initial polarized state (shown in **Fig. 5**), that indicates the existence of a depolarized state (shown in **Fig. 6**). As for the initial state, similar $U_{\pi 1,2}(f_m)$, $\varphi_{\pi 1,2}(f_m)$ and $D_\pi(f_m)$ dependences and near equal $U_{\pi 1,2}$ values are observed under probing opposite surfaces of the sample. Together with this, the decrease of $U_{\pi 1}$ and $f_m U_{\pi 2}$ values with increase in $f_m$ is stronger than for the initial state and corresponds to about 10 times decrease of $\gamma$ and $\gamma/\varepsilon_\pi$ values from the volume to the surface. The difference between the values of $U_{\pi 1,2}$ and $f_m U_{\pi 2}$ for opposite surfaces of the partially depolarized sample increases under the transition from the volume to the surface probing and reaches 2 – 3 times for $f_m = 400$ Hz (that corresponds to $\lambda_T =$ 15 μm). The increase of $\varphi_{\pi 1,2}$with the increase in $f_m$ can be associated with the delay of the heat transfer through the subsurface layer with decreased pyroelectric activity. The peculiarities of $f_m U_{\pi 2}$ and $\varphi_{\pi 1,2}$ in the vicinity of $f_m \approx 200$ Hz (see left side of **Fig. 6**) and $f_m \approx 500$ Hz (see right side of **Fig. 6**) indicate some under-surface inhomogeneities of $\gamma/\varepsilon_\pi$ distribution located under the surfaces of the sample at the depth corresponding to $\lambda_T \approx 25$ μm for the left side of **Fig. 6** and $\lambda_T \approx 15$ μm for the right side of **Fig. 6**.

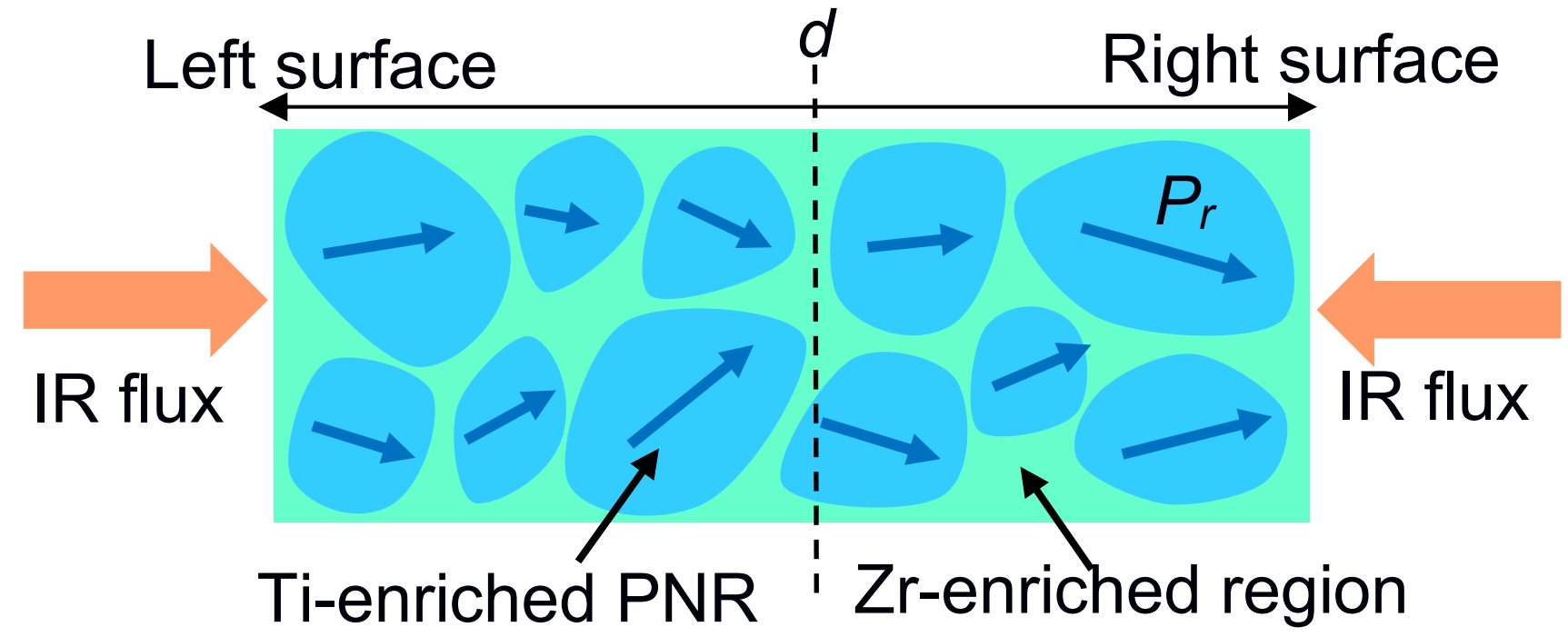

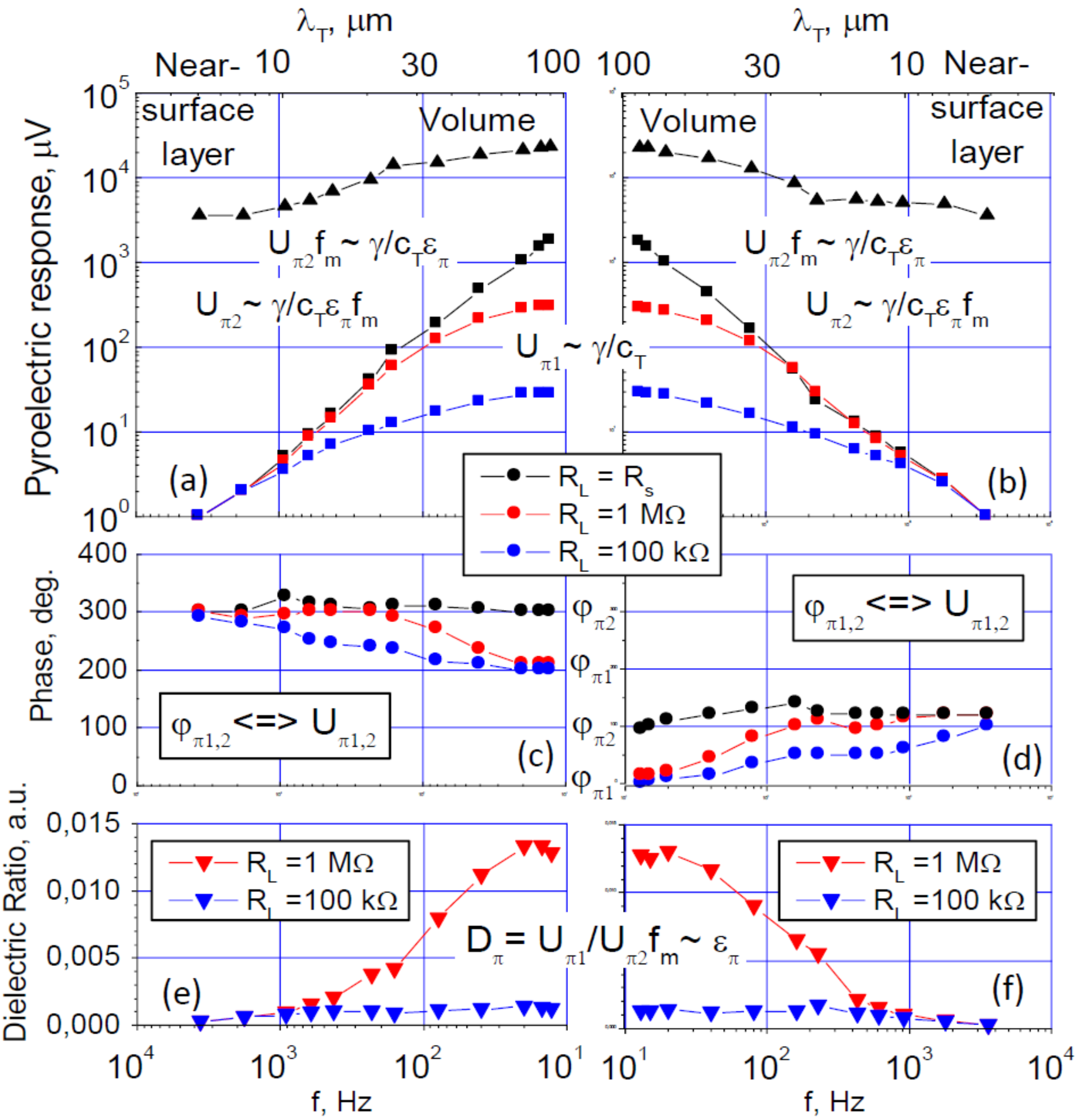

**FIGURE 5. Initial polarized state of as prepared BZT-20 ceramic pellet is probed:** blue arrows, which are mainly co-directed, point the direction of the remanent polarization $P_r$ in the PNRs (see the top scheme)**.** Dependences of pyroelectric response amplitude **(a, b)**, phase **(c, d)** and dielectric ratio **(e, f)** on the IR-flux modulation frequency $f_m$ (see lower scale) and thermal diffusion length $\lambda_T$ (see upper scale). The cases in which either the left surface **(a, c, e**) or the right surface **(b, d, f)** of the pellet is IR-irradiated, are shown by the left and right columns, respectively.

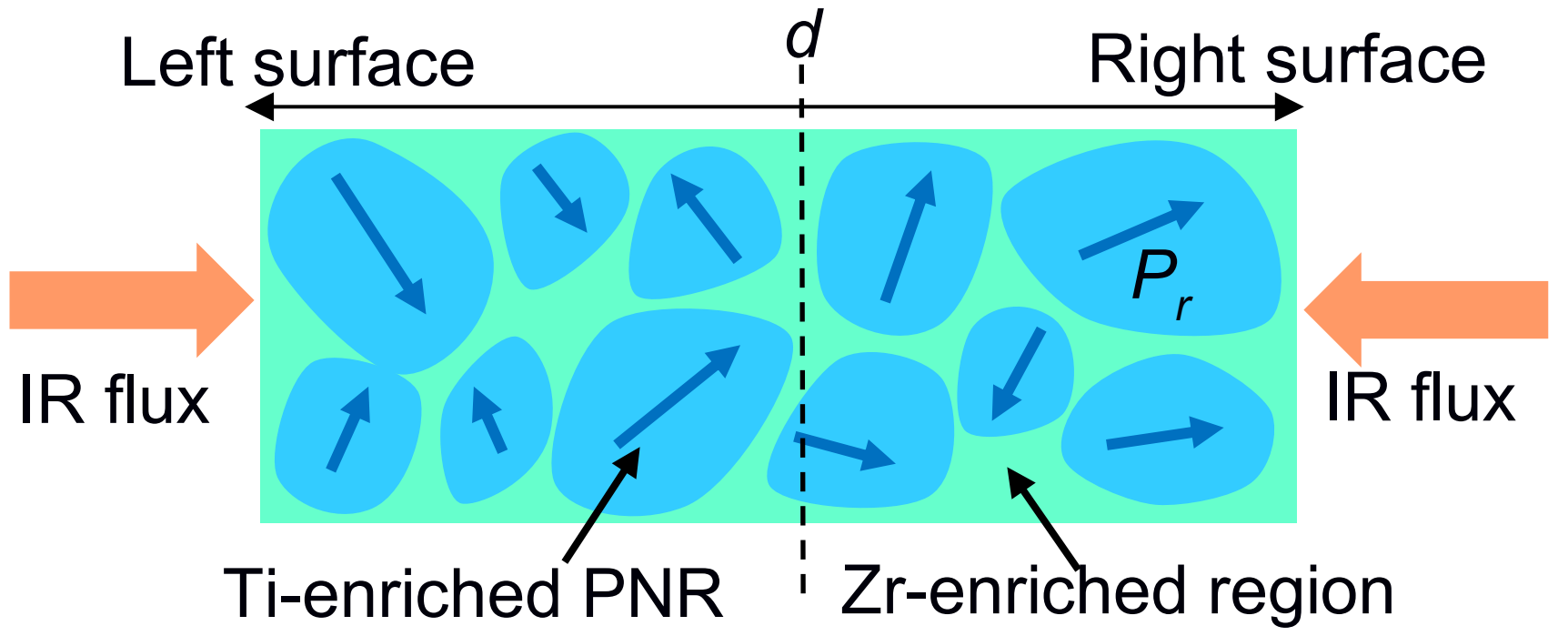

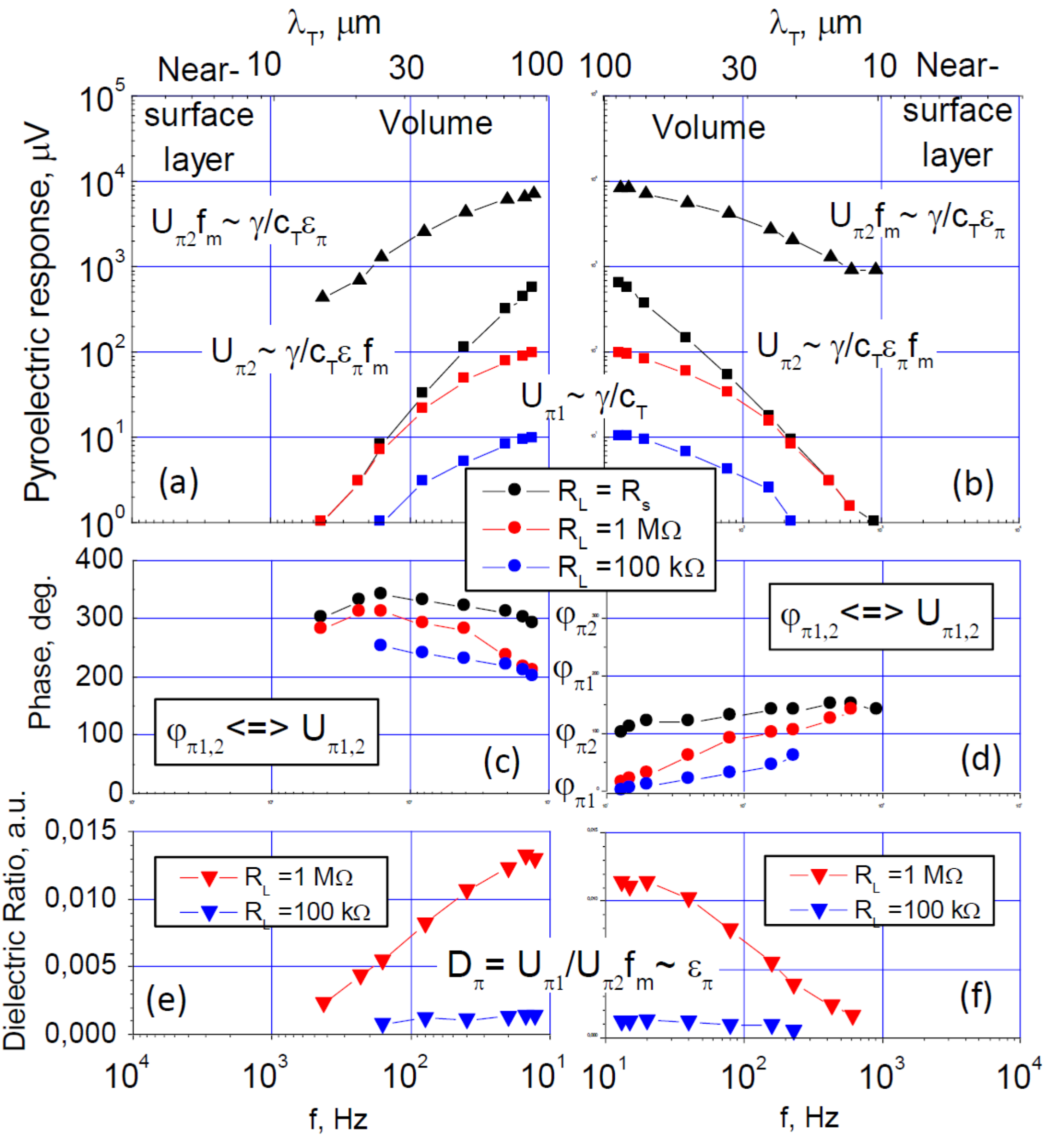

**FIGURE 6. An almost symmetric partially depolarized state of BZT-20 ceramic pellet is probed:** blue arrows, which are partially misaligned near both surfaces, point the direction of the remanent polarization $P_r$ in the PNRs (see the top scheme). Dependences of the pyroelectric response amplitude **(a, b)**, phase **(c, d)** and dielectric ratio **(e, f)** on the IR-flux modulation frequency $f_m$ (lower scale) and thermal diffusion length $\lambda_T$ (upper scale). The cases in which either the left surface **(a, c, e)** or the right surface **(b, d, f)** of the pellet is IR-

irradiated, are shown by the left and right columns, respectively. The partially depolarized state is formed after the application of 10 Hz sine voltage with amplitude ±300 V and its subsequent gradual vanishing to zero voltage.

It appeared that partially depolarized states can be made asymmetric with respect to the opposite surfaces of the sample after application of higher periodic voltage at higher frequency. For instance, a *strongly asymmetric depolarized state,* hereinafter called "*semi-depolarized*", was formed after the application of a 20 Hz sine voltage with an amplitude of ±400 V and its subsequent gradual vanishing to zero. Pyroelectric response of the semi-depolarized state of BZT-20 ceramic pellet is shown in **Fig. 7**. For the state a significant difference of $U_{\pi 1,2}$ values and different form of $f_m U_{\pi 2}$ and $U_{\pi 2}$ frequency dependences under probing of opposite surfaces are observed. The decrease of $f_m U_{\pi 2}$ values with $f_m$ increase is stronger for the left surface of the sample, than that for the right one, which corresponds to the different behavior of the $\gamma/\varepsilon_\pi$ values from volume to the surface for the opposite surfaces. These features indeed confirm the existence of a so called "semi-depolarized" state.

A comparison of the dependences $\varphi_{\pi 1,2}(f_m)$ to the initial state and partially depolarized state shows the similar behavior of $\varphi_{\pi 1,2}(f_m)$ for the right surface and the difference in $\varphi_{\pi 1,2}(f_m)$ behavior with noticeable difference of $\varphi_{\pi 1,2}$ values at low frequences $f_m$ (i.e., for $\lambda_T > 30$ μm) for the left surface. Such asymmetry of $f_m U_{\pi 2}$ and $U_{\pi 1,2}$ frequency dependences, as well as the change of $\varphi_{\pi 1,2}(f_m)$ behavior, is the sign of the existence of a depolarized region with the decreased $\gamma$ and $\gamma/\varepsilon_\pi$ values near the corresponding surface of the sample. When probing opposite surfaces of the sample, the phase $\varphi_{\pi 1,2}(f_m)$ increases with the increasing $f_m$, which reflects the existence of a pyroelectric inhomogeneity in the sample. It should be noted that the increase in $\varphi_{\pi 1,2}(f_m)$ can also be associated with the subsurface thermal inhomogeneity of the sample. It is known that in the case of a two-layer thermal inhomogeneity, the function $\varphi_{\pi 1,2}(f_m)$ should increase linearly with $\sqrt{f_m}$ due to the dependence of the phase shift $\Delta\varphi$ on the thickness $d$ of the nonpolar gap: $\Delta\varphi = d/\lambda_T = d\sqrt{\pi f_m/a_T} \sim \sqrt{f_m}$ [54, 55]. It is likely that the observed nearly logarithmic increase of $\varphi_{\pi 1,2}(f_m)$ with frequency can be attributed to both thermal and pyroelectric inhomogeneity of the sample.

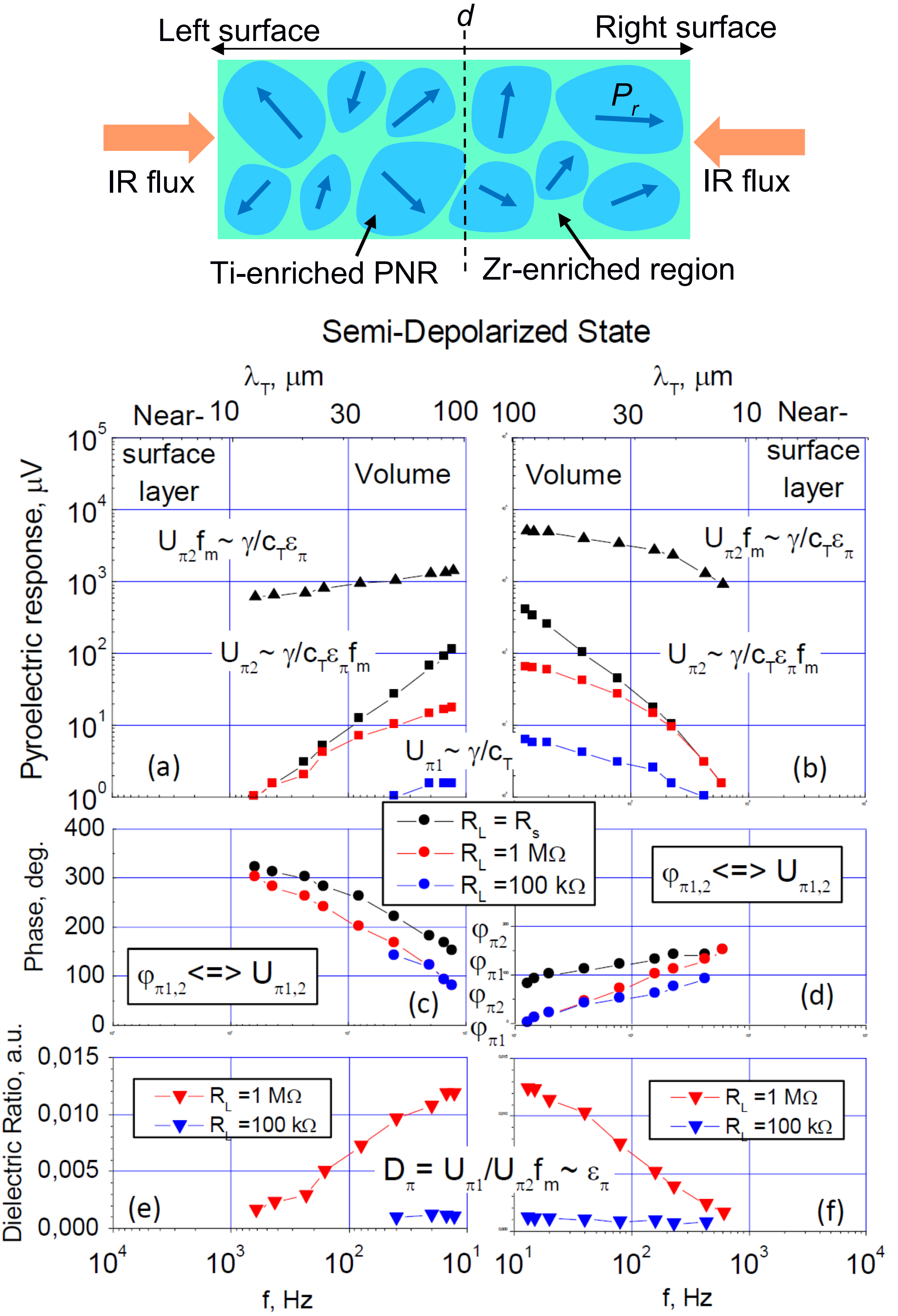

**FIGURE 7. An asymmetric semi-depolarized state of BZT-20 ceramic pellet is probed:** blue arrows, which are completely misaligned near the left surface and partially aligned near the right surface, point the direction of the remanent polarization $P_r$ in the PNRs (see the top scheme). Dependences of the pyroelectric response amplitude **(a, b)**, phase **(c, d)** and dielectric ratio **(e, f)** on the IR-flux modulation frequency $f_m$ (lower scale) and thermal diffusion length $\lambda_T$ (upper scale). The cases in which either the left surface **(a, c, e)** or the right surface **(b, d, f)** of the pellet is IR-irradiated, are shown by the left and right columns, respectively. The asymmetric semi-depolarized state is formed from the partially depolarized after the application of a 20 Hz sine voltage with an amplitude of ±400 V and its subsequent gradual vanishing to zero.

The electric field-induced asymmetric semi-depolarized state is unstable and has the tendency to return gradually to a symmetric state. After a 24-hours (or longer) relaxation, more symmetrical $f_m U_{\pi 2}(f_m)$ and $U_{\pi 1,2}(f_m)$ dependences, as well as $\varphi_{\pi 1,2}(f_m)$ dependences, which are inherent to the partially depolarized state, are observed. A typical pyroelectric response of the *relaxed depolarized state* of BZT-20 ceramic pellet is shown in **Fig. 8**. However, the $U_{\pi 1,2}$ and $U_{\pi 2} f_m$ values, as well as the values of $\gamma$ and $\gamma/\varepsilon_\pi$, remain considerably smaller than those observed in the initial state and those of the partially depolarized state. The pyroelectric coefficient $\gamma$, estimated in the relaxed disordered state, is about $(0.5 - 5)10^{-3}$ $C/m^2K$ depending on the frequency $f_m$, that is significantly higher than the $\gamma$ values characteristic for BST ceramics [45, 46]. The 10-times increase of $\gamma$ probably originated from the relaxor ferroelectric state.

The *"repolarized" state* with completely reversed polarization was achieved by application of –400 V, 20 Hz semi-sine voltage. Results of the pyroelectric probing of the repolarized sample of BZT-20 ceramics are shown in **Fig. 9**. This state looks like the initially polarized one, but with the inverted $\varphi_{\pi 1,2}(f_m)$ dependences (left and right sides), which confirms the fact of the polarization reversal. Obtained values of $U_{\pi 2}\, f_m$ and $U_{\pi 1,2}$, and the values of $\gamma$ and $\gamma/\varepsilon_\pi$, are close to those inherent to the initially polarized state, however the phases $\varphi_{\pi 1,2}$ are shifted at approximately 180°C (compare **Fig. 9** and **Fig. 5**). The peculiarity of $f_m U_{\pi 2}(f_m)$ in the vicinity of $f_m \approx 2$ kHz corresponds to some subsurface inhomogeneity of $\gamma/\varepsilon_\pi$ located under the surface of the sample at the depth corresponding to $\lambda_T \approx 5$ µm.

The results of the pyroelectric probing of the BZT-20 sample with fully *restored polarization* is shown in **Fig. 10.** The restored polarized state was achieved by application of +400 V, 20 Hz semi-sine voltage. In this way the polarized state, similar to the initial one, was observed. The dependences $f_m U_{\pi 2}(f_m)$, $U_{\pi 1,2}(f_m)$ and $\varphi_{\pi 1,2}(f_m)$ have the same form as in the initial state, except for the peculiarities in the vicinity of $f_m \approx 200$ Hz. Obtained values of $U_{\pi 1,2}$ and $U_{\pi 2}\, f_m$, as well as the values of $\gamma$ and $\gamma/\varepsilon_\pi$, are close to those inherent to the initial polarized state. The polarized state is more uniform than the initial one at the right surface; it is similar to the initial one at the left surface. Thus, we can conclude that electric field cycling improves the ferroelectric polarization and pyroelectric response of the relaxor BZT-20 ceramics.

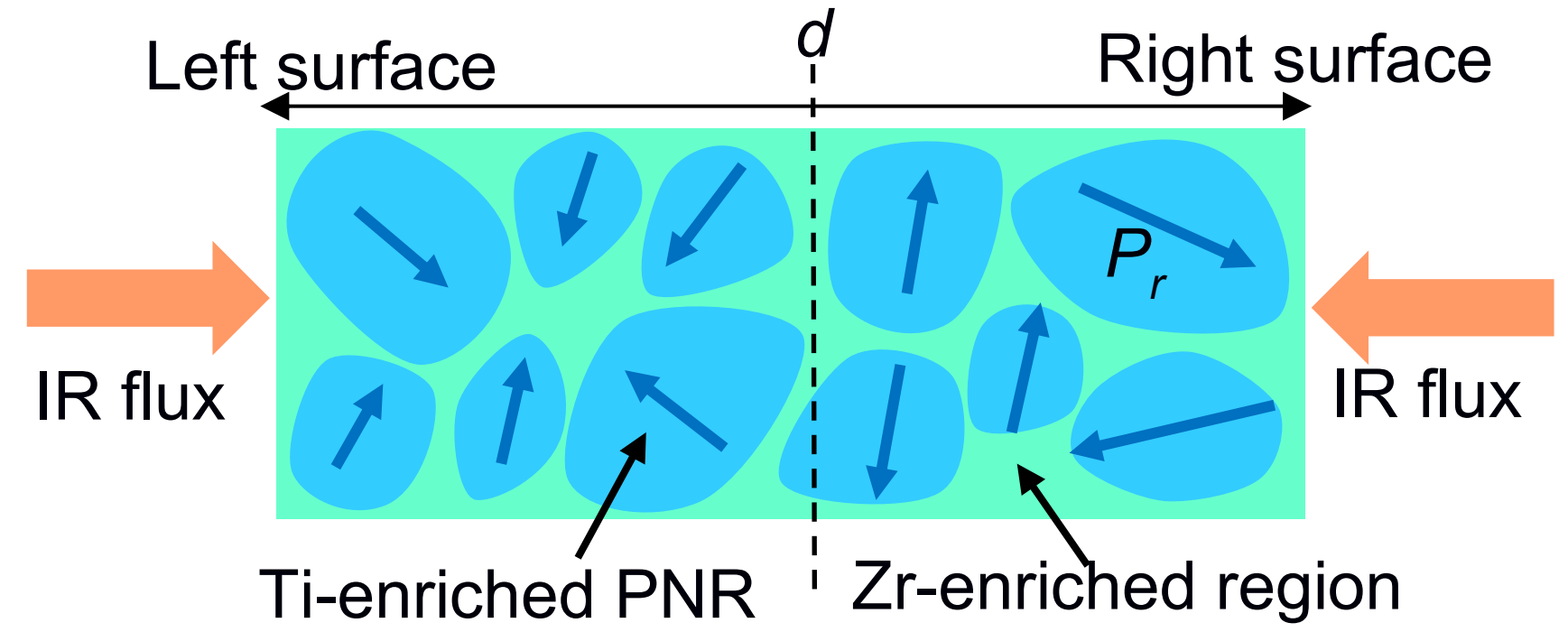

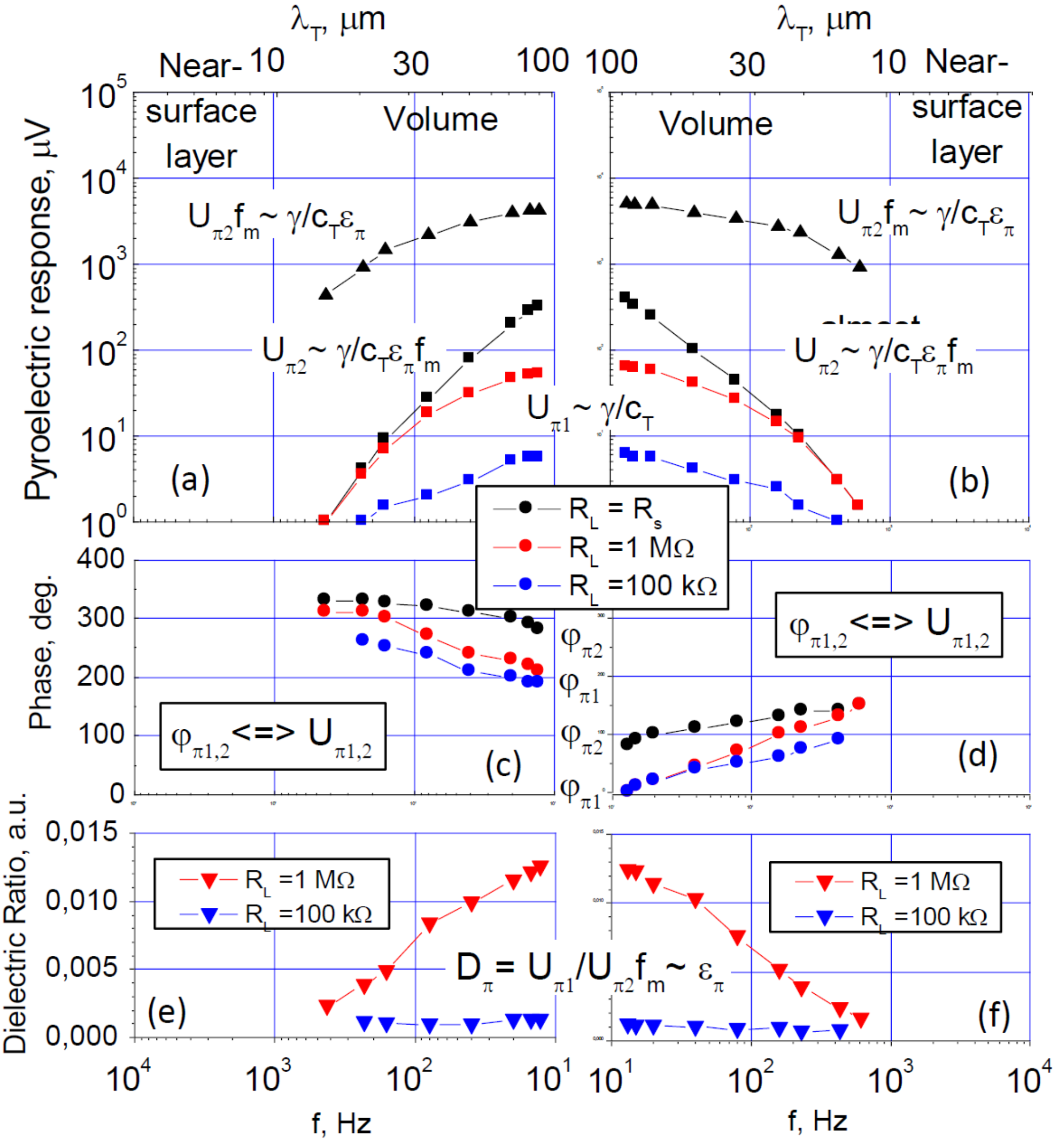

**FIGURE 8. A relaxed depolarized state of BZT-20 ceramic pellet is probed:** blue arrows, which are completely misaligned, point the direction of the remanent polarization $P_r$ in the PNRs (see the top scheme). Dependences of the pyroelectric response amplitude **(a, b)**, phase **(c, d)** and dielectric ratio **(e, f)** on the IR-flux modulation frequency $f_m$ (lower scale) and thermal diffusion length $\lambda_T$ (upper scale). The cases in which either the left surface **(a, c, e)** or the right surface **(b, d, f)** of the pellet is IR-irradiated, are shown by the left

and right columns, respectively. The state is formed spontaneously after 24 hours (or longer) relaxation of the asymmetric semi-polarized state.

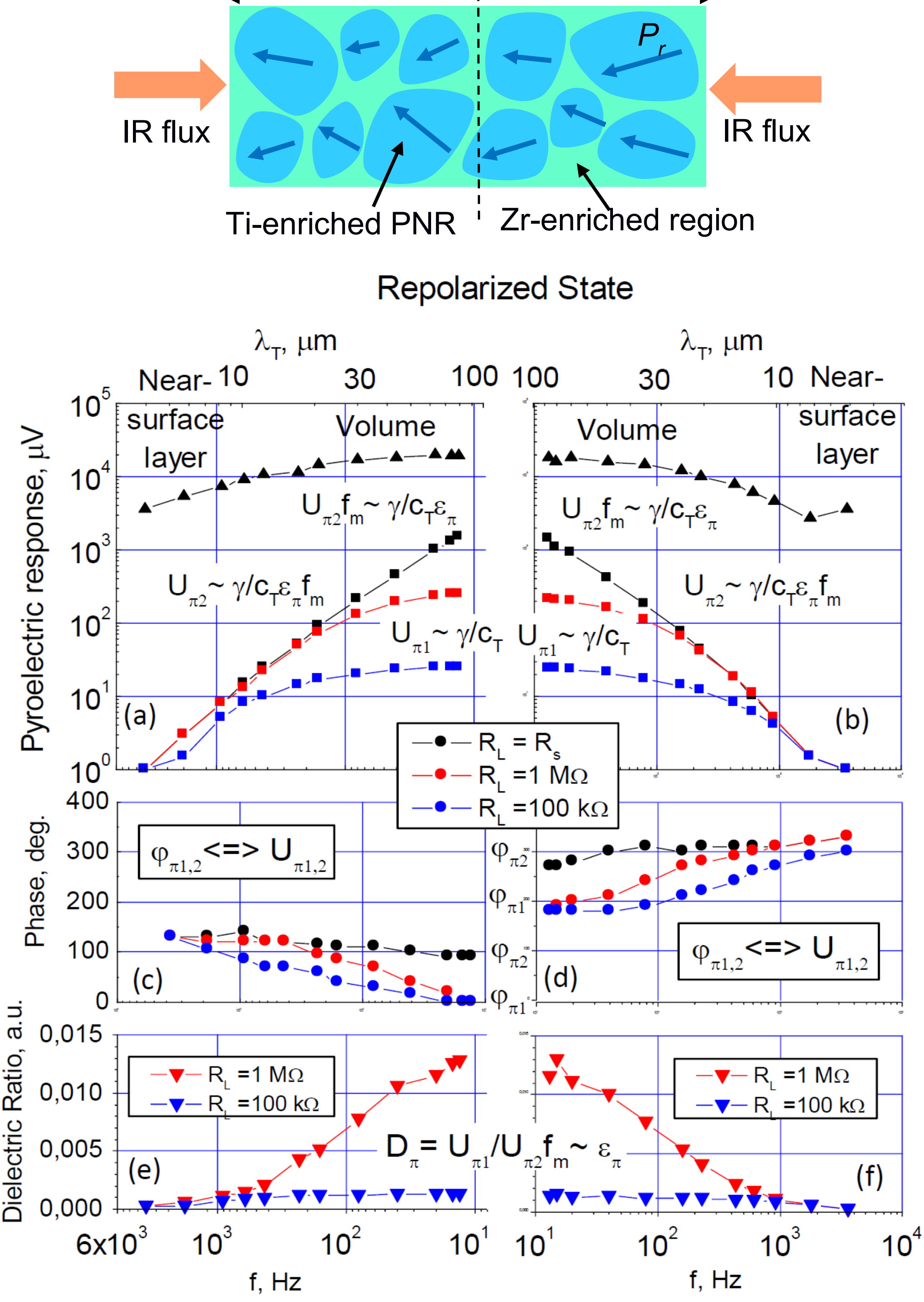

**FIGURE 9. A repolarized state of BZT-20 ceramic pellet is probed:** blue arrows, which are mainly co-directed, point the direction of the remanent polarization $P_r$ in the PNRs (see the top scheme)**.** The averaged $P_r$ is opposite to those shown in Fig. 5. Dependences of pyroelectric response amplitude **(a, b)**, phase **(c, d)**

and dielectric ratio **(e, f)** on the IR-flux modulation frequency $f_m$ (lower scale) and thermal diffusion length $\lambda_T$ (upper scale). The cases in which either the left surface **(a, c, e)** or the right surface **(b, d, f)** of the pellet is IR-irradiated, are shown by the left and right columns, respectively. The repolarized state with completely reversed polarization is formed by application of –400 V, 20 Hz semi-sine voltage.

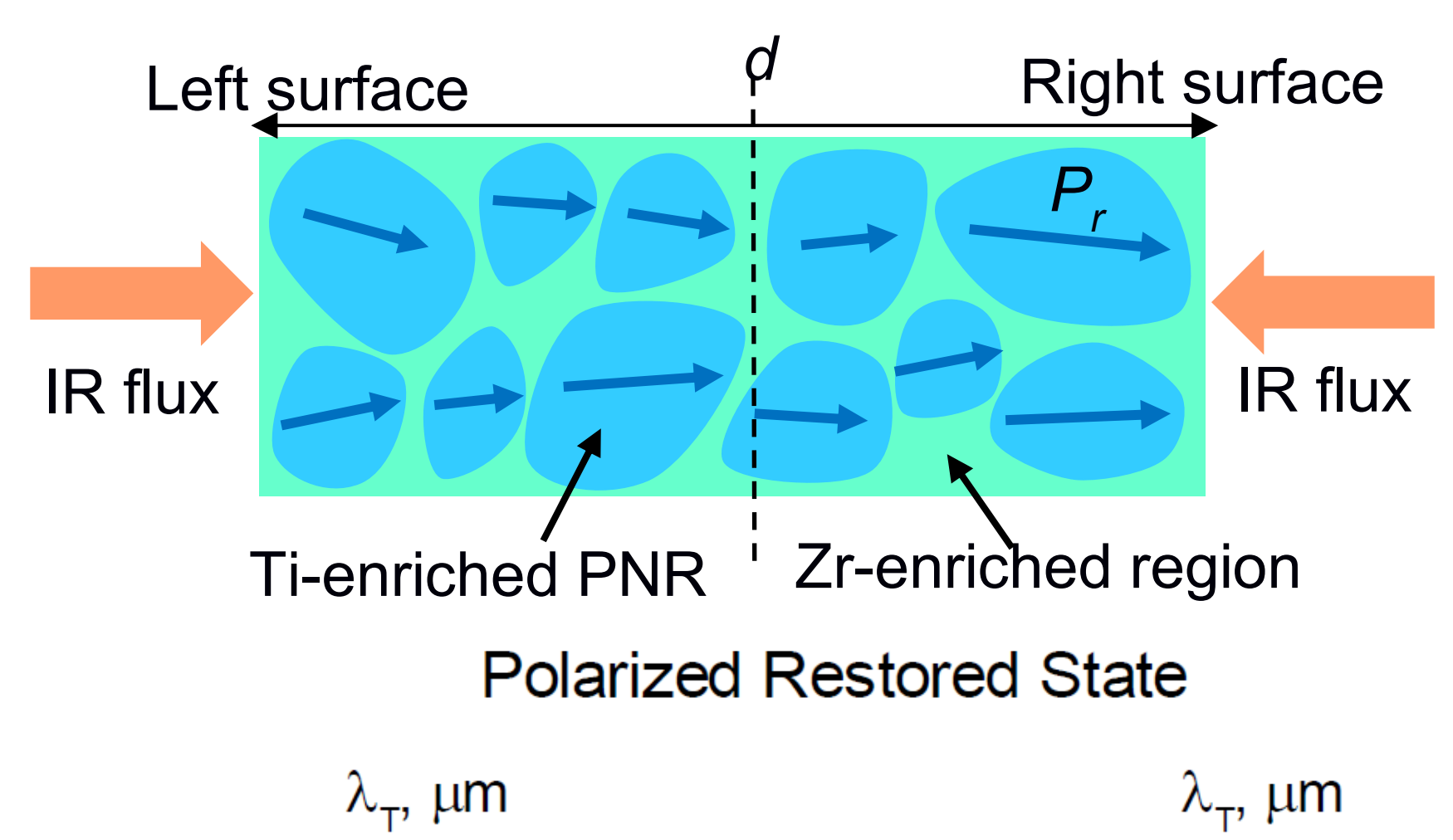

**FIGURE 10. A restored polarized state of BZT-20 ceramic pellet is probed**: blue arrows, which are almost

co-directed, point the direction of the remanent polarization $P_r$ in the PNRs (see the top scheme). Dependence of pyroelectric response amplitude **(a, b)**, phase **(c, d)** and dielectric ratio **(e, f)** on the IR-flux modulation frequency $f_m$ (lower scale) and thermal diffusion length $\lambda_T$ (upper scale). The cases in which either the left surface **(a, c, e**) or the right surface **(b, d, f)** of the pellet is IR-irradiated, are shown by the left and right columns, respectively. The restored polarized state was achieved by application of +400 V, 20 Hz semi-sine voltage.

The relaxor-type temperature decay of the remanent polarization $P_r$ inside individual PNRs can be approximated by the expression [56]:

$$P_r \approx \begin{cases} P_0, & T < T_f, \\ P_0\left[1 - \exp\left(-\frac{T_a}{T-T_f}\right)\right], & T > T_f. \end{cases} \tag{2a}$$

$$\Pi = -\Sigma \approx \begin{cases} 0, & T < T_f, \\ \Pi_0 \exp\left(-\frac{T_a}{T-T_f}\right)\frac{T_a^2}{(T-T_f)^2}, & T > T_f, \end{cases} \tag{2b}$$

where $T_a$ is the activation energy of atomic thermal fluctuations (in temperature units), which suppress and eventually destroy the long-range polar order; $T_f$ is the "local freezing" temperature of electric dipolar order, that is primarily determined by the local content x of Zr ions, as well as by concentration of various defects (e.g., oxygen vacancies). The exponential function, $\exp\left(-\frac{T_a}{T-T_f}\right)$, corresponds to the Vogel-Fulcher law valid for relaxor ferroelectrics [57] and dipolar glasses [58, 59]. The parameter $\Pi_0 = \frac{P_0}{T_a}$, introduced in Eq.(2b), is the magnitude of the pyroelectric/electrocaloric effect.

Expressions (2) should be averaged over the random orientation of the crystallographic axes inside different grains, but the averaging should consider the long-range interactions between PNRs, which can be located either inside microsized grains, or in neighboring grains of BZT-20 ceramics. Let us assume that the polarization $\vec{P}(\vec{E})$ satisfies the Landau-Ginzburg-Devonshire-type equation inside individual PNRs. This is a rough approximation; however it allows us to explain experimental trends. As a result of statistical averaging, that includes also the averaging over PNR sizes, we obtained that the pyroelectric and electrocaloric coefficients of the BZT-20 ceramics in the relaxor state at $T < \langle T_f \rangle$, as well as in the mixed ferroelectric-relaxor state, is much larger than those in the ordered ferroelectric state. The difference is reminiscent with the strong increase of local dielectric permittivity in the ferroelectric-relaxor state.

## 4. CONCLUSIONS

$Ba_xZr_{1-x}TiO_3$ ceramics with $0.15 < x < 0.25$, which have high dielectric permittivity, small leakage currents and low dissipation factor, are promising materials for tunable capacitor devices, multilayered ceramic capacitors and piezoelectric actuators. These materials reveal relaxor properties due to the partial isovalent substitution of $Ti^{4+}$ ions by $Zr^{4+}$ ions with a larger ionic size, which leads to inhomogeneous local chemical strains, whose in turn diffuse and broaden the ferroelectric-paraelectric transition. However, to the best of our knowledge, the pyroelectric properties of $Ba_xZr_{1-x}TiO_3$ ceramics with x~0.2 have not been studied. To fill the gap in knowledge, we perform the pyroelectric thermo-wave probing of $BaZr_{0.2}Ti_{0.8}O_3$ ceramics prepared by the solid-state synthesis.

Results of the thermo-wave probing of the most interesting polar states, namely polarized, depolarized, relaxed and restored polarized states, reveal the pronounced pyroelectric response that can be strongly asymmetric with respect to the opposite surfaces of the ceramics. The asymmetry and profiles of pyroelectric response depend significantly on the pre-history of the electric field cycling indicating possible non-ergodic relaxor-type polar states in the ceramic sample. Interesting, that electric field cycling improves the ferroelectric polarization and pyroelectric response of the relaxor BZT-20 ceramics.

X-ray diffraction spectrum, recorded at room temperature, reveals the virtual absence of the macroscopic tetragonality and the weak asymmetry of the (200) peak, which indicate the small tetragonality inside the Ti-enriched polar nanoregions. Decrease of the relative intensity of the Raman band at 714 $cm^{-1}$ occurring upon heating evidences the diffuse ferroelectric-paraelectric phase transition between 30°C and 40°C. The temperature dependence of the dielectric permittivity obeys modified Curie-Weiss law with power 1.4, indicating that both ordered ferroelectric and relaxor states coexist in the $BaTi_{0.8}Zr_{0.2}O_3$ ceramics.

Estimated pyroelectric coefficients, which are $\sim 5\cdot 10^{-4}$ $C/m^2K$ in the ordered ferroelectric state and at least 10 times larger in the relaxor ferroelectric state, are relatively high. Thus, obtained results can be useful for elaboration of lead-free relaxor ferroelectric ceramics for pyroelectric and electrocaloric applications.

**Authors' contribution**. N.V.M. and M.Y. performed the measurements of pyroelectric response and analyzed the results obtained. Y.O.Z. and L.P.Y. performed measurements of polarization hysteresis and dielectric permittivity; and analyzed the results jointly with O.S.P, who also prepared samples for the pyroelectric measurements and processed the data. V.B. and V.V.L. sintered the $Ba_xZr_{1-x}TiO_2$ ceramics. V.M.D. performed SEM, EDS, XRD and Raman spectroscopy studies and analyzed the results obtained jointly with M.O.S. I.V.K. provided microscopic explanation of the relaxor properties of $Ba_xZr_{1-x}TiO_2$ ceramics. E.A.E. and A.N.M. elaborated theoretical interpretation of the experimental results and prepared corresponding figures, wrote introduction and made conclusions. Corresponding authors (E.A.E., V.O.D. and A.N.M.) made all improvements in the manuscript.

**Acknowledgments.** N.V.M., M.Y., O.S.P., M.O.S. and A.N.M. acknowledge the Target Program of the National Academy of Sciences of Ukraine, Projects No. 5.8/26-П "Energy-saving and environmentally friendly nanoscale ferroics for the development of sensorics, nanoelectronics and spintronics". Y.O.Z., L.P.Y., I.V.K. and E.A.E. acknowledge the support of the National Academy of Sciences of Ukraine (project "Relaxor ferroelectrics with perovskite structure for capacitors, actuators, pyroelectric sensors, and electrocaloric coolers"). V.M.D. work is funded by the National Research Foundation of Ukraine (grant N 2023.05/0022 "Establishment of the State Key Laboratory «Center for Critical Optoelectronic Micro- and Nano- Technologies and Expertise»"). A.N.M. and V.V.L. acknowledge the NATO Science for Peace and Security Programme under grant SPS G5980 "FRAPCOM" for sponsoring ceramics synthesis.

## REFERENCES

[1] B. Jaffe, W.R. Cook, Jr., and H.L. Jaffe, "Piezoelectric Ceramics". Academic Press, London & New York (1971).

[2] G.H. Haertling, "Ferroelectric ceramics: history and technology", J. Am. Ceram. Soc. **82**, 797-818 (1999);

[3] M. M. Kržmanc, H. Uršič, A. Meden, R. C. Korošec, and D. Suvorov. "$Ba_{1-x}Sr_xTiO_3$ plates: synthesis through topochemical conversion, piezoelectric and ferroelectric characteristics." Ceramics International **44**, 21406 (2018). https://doi.org/10.1016/j.ceramint.2018.08.198

[4] L. S. Kremenchugsky, "Segnetoelectricheskije Prijomniki Izluchenija", (Ferroelectric Detectors of Radiation). Naukova Dumka, Kiev, (1971) (in Russian).

[5] S.B. Lang, "Sourcebook of Pyroelectricity", Gordon & Breach Sci. Publ, London - NewYork - Paris, (1974).

[6] S.B. Lang, "Pyroelectricity: From ancient curiosity to modern imaging tool", Physics Today, **58**(8), 31-36 (2005); https://doi.org/10.1063/1.2062916

[7] D. Zhang, H. Wu, C.R. Bowen, Y. Yang, "Recent advances in pyroelectric materials and applications", Small, **17**(51), e2103960 (2021); https://doi.org/10.1002/smll.202103960

[8] G. Zhang, X. Zhang, T. Yang, Qi Li, L.-Q. Chen, S. Jiang, and Q. Wang. Colossal room-temperature electrocaloric effect in ferroelectric polymer nanocomposites using nanostructured barium strontium titanates. ACS nano **9**, 7164 (2015); https://doi.org/10.1021/acsnano.5b03371

[9] Z. Hanani, D. Mezzane, M. Amjoud, M. Lahcini, M. Spreitzer, D. Vengust, A. Jamali, M. El Marssi, Z. Kutnjak, and M. Gouné. "The paradigm of the filler's dielectric permittivity and aspect ratio in high-k

polymer nanocomposites for energy storage applications." Journal of Materials Chemistry **C 10**, 10823 (2022); https://doi.org/10.1039/d2tc00251e

[10] A.N. Morozovska, O. S. Pylypchuk, S. Ivanchenko, E. A. Eliseev, H. V. Shevliakova, L. M. Korolevich, L. P. Yurchenko, O. V. Shyrokov, N. V. Morozovsky, V. N. Poroshin, Z. Kutnjak, and V. V. Vainberg. Size-induced High Electrocaloric Response of the Dense Ferroelectric Nanocomposites. Ceramics International **50** (7b), 11743 (2024); https://doi.org/10.1016/j.ceramint.2024.01.079

[11] Z. Hanani, D. Mezzane, M. Amjoud, M. Lahcini, M. Spreitzer, D. Vengust, A. Jamali, M. El Marssi, Z. Kutnjak, and M. Gouné. "The paradigm of the filler's dielectric permittivity and aspect ratio in high-k polymer nanocomposites for energy storage applications." Journal of Materials Chemistry **C 10**, 10823 (2022); https://doi.org/10.1039/d2tc00251e

[12] M. Kumari, M. Chahar, S. Shankar, and O. P. Thakur. "Temperature dependent dielectric, ferroelectric and energy storage properties in Bi0.5Na0.5TiO3 (BNT) nanoparticles." Materials Today: Proceedings **67**, 688 (2022); https://doi.org/10.1016/j.matpr.2022.06.542

[13] Z. Luo, Z. Ye, B. Duan, G. Li, K. Li, Z. Yang, S. Nie, T. Chen, L. Zhou, and P. Zhai. "SiC@ BaTiO3 core-shell fillers improved high temperature energy storage density of P (VDF-HFP) based nanocomposites." Composites Science and Technology **229**, 109658 (2022), https://doi.org/10.1016/j.compscitech.2022.109658

[14] Y. Sakabe, N. Wada, T. Hiramatsu, and T. Tonogaki, "Dielectric properties of fine-grained BaTiO3 ceramics doped with CaO". Jpn. J. Appl. Phys. **41**(5A), 6922 (2002); https://doi.org/10.1143/JJAP.41.6922

[15] D.-H. Yoon, "Tetragonality of barium titanate powder for a ceramic capacitor application", Journal of Ceramic Processing Research., **7**(4), 343-354 (2006);

[16] D. Hennings, A. Schnell, G. Simon, "Diffuse ferroelectric phase transitions in Ba(Ti1-yZry)O3 ceramics". J. Am. Ceram. Soc., **65**(11), 539-544 (1982); https://doi.org/10.1111/j.1151-2916.1982.tb10778.x

[17] H. Kishi, Y. Mizuno, H. Chazono, "Base-metal electrode-multilayer ceramic capacitors: Past, present and future perspectives". Jpn. J. Appl. Phys., **42**(1R), 1-15 (2003); http://doi.org/10.1143/JJAP.42.1

[18] X.G. Tang, K.H. Chen, and H.L.W. Chan, "Diffuse phase transition and dielectric tunability of Ba(ZryTi1−y)O3 relaxor ferroelectric ceramics". Acta Mater., **52**(17), 5177-5183 (2004); https://doi.org/10.1016/j.actamat.2004.07.028

[19] R.D. Shannon, "Revised effective ionic radii and systematic studies of interatomic distances in halides and chalcogenides," Acta Crystallographica A, **32**(5) 751-767 (1976); https://doi.org/10.1107/S0567739476001551

[20] V.V. Shvartsman, and D.C. Lupascu, "Lead-free relaxor ferroelectrics", J. Am. Ceram. Soc. **95**, 1-26 (2012); https://doi.org/10.1111/j.1551-2916.2011.04952.x

[21] W. Li, Z Xu, R Chu, P Fu, and J Hao, "Sol–gel synthesis and characterization of Ba(1−x)SrxTiO3 ceramics", Journal of Alloys and Compounds, **499**, 255-258 (2010); https://doi.org/10.1016/j.jallcom.2010.03.180

[22] K. Zhang, Y. Xu, L. Chen, L. Fu, S. Jia, Le Cao and Q. Zhang, “Effect of Zr4+ content on crystal structure, micromorphology and dielectric properties of Ba(ZrxTi1-x)O3 ceramics”. J. Appl. Mat. Sci. & Eng. Res. **3**, 5 (2019); https://doi.org/10.33140/amse.03.01.1

[23] H. Wen, X. H. Wang, L. T. Li, & Z. L. Gui, “Properties of Sr-doped BaTiO3 based X7R ceramic materials”. Key Engineering Materials, **280**, 65-68 (2004); https://doi.org/10.4028/www.scientific.net/KEM.280-283.65

[24] C. Fu, F. Pan, H. Chen, C. Wei, and C. Yang, “Effect of annealing on leakage current characteristics of Pt/Ba0.6Sr0.4TiO3/Pt thin-film capacitors”. J. Mater. Sci.: Mater. Electron, **18**, 453-456 (2007); https://doi.org/10.1007/s10854-006-9054-y

[25] B. Ruthramurthy, K. Gebremedhn Kelele, H. A. Murthy, K. B. Tan, K.Y. Chan, D. Muniswamy, A. Tadesse, & S. Ghotekar, “Multielement doped barium strontium titanate nanomaterials as capacitors”. Journal of Chemistry, **2023**, 6338649 (2023); https://doi.org/10.1155/2023/6338649

[26] A.K. Tagantsev, V.O. Sherman, K.F. Astafiev, J. Venkatesh, N. Setter, “Ferroelectric materials for microwave tunable applications”. Journal of Electroceramics, **11**, 5-66 (2003); https://doi.org/10.1023/B:JECR.0000015661.81386.e6

[27] V.R. Mudinepalli, L. Feng, W.C. Lin, and B. S. Murty, “Effect of grain size on dielectric and ferroelectric properties of nanostructured Ba0.8Sr0.2TiO3 ceramics”, J. Adv. Ceram. **4**, 46-53 (2015); https://doi.org/10.1007/s40145-015-0130-8

[28] X. Jili, L. Gao, H. Chen, J. Zhang, “Microstructure and dielectric properties of gradient composite BaxSr1−xTiO3 multilayer ceramic capacitors”. Micromachines **15**, 470-477 (2024); https://doi.org/10.3390/mi15040470

[29] U. Weber, G. Greuel, U. Boettger, S. Weber, D. Hennings, & R. Waser, “Dielectric properties of Ba(Zr, Ti)O3-based ferroelectrics for capacitor applications”. Journal of the American Ceramic Society, **84**, 759-766 (2001); https://doi.org/10.1111/j.1151-2916.2001.tb00738.x

[30] T. Hoshina, T. Furuta, T. Yamazaki, H. Takeda, and T. Tsurumi, “Grain size effect on dielectric properties of Ba(Zr,Ti)O3 ceramics”, Jpn. J. Appl. Phys., **51**, 09LC04 (2012); http://doi.org/10.1143/JJAP.51.09LC04

[31] S.-H. Yoon, J.-R. Kim, S.-H. Yoon, C.-H. Kim, and D.-Y. Kim, “Resistance degradation behavior of Zr-doped BaTiO3 ceramics and multilayer ceramic capacitor”, J. Mater. Res. **28**, 1078-1086 (2013); http://doi.org/10.1557/jmr.2013.57

[32] Z. Yu, C. Ang, R. Guo and A.S. Bhalla, “Dielectric properties and high tunability of BaTi0.7Zr0.3O3 ceramics under dc electric field”, Appl. Phys. Lett. **81**, 1285-1287 (2002); http://doi.org/10.1063/1.1498496

[33] Q. Xu, D. Zhan, D.-P. Huang, H.-X. Liu, W. Chen, F. Zhang, “Dielectric inspection of BaZr0.2Ti0.8O3 ceramics under bias electric field: A survey of polar nano-regions”. Materials Research Bulletin, **47**, 1674-1679 (2012); http://doi.org/10.1016/j.materresbull.2012.03.062

[34] P. W. Rehrig, S. E. Park, S. Trolier-McKinstry, G. L. Messing, B. Jones, & T. R. Shrout, "Piezoelectric properties of zirconium-doped barium titanate single crystals grown by templated grain growth". J. Appl. Phys. **86**, 1657-1661 (1999); https://doi.org/10.1063/1.370943

[35] Z. Yu, C. Ang, R. Guo, A.S. Bhalla, "Piezoelectric and strain properties of Ba(Ti1-xZrx)O3 ceramics". J. Appl. Phys. **92**, 1489-1493 (2002); https://doi.org/10.1063/1.1487435

[36] Z. Yu, R. Guo, and A. S. Bhalla, "Dielectric behavior of Ba(Ti1-xZrx)O3 single crystals" J. Appl. Phys., **88**, 410 (2000); http://doi.org/10.1063/1.373674

[37] H. Chen, C. Yang, C. Fu, J. Shi, J. Zhang, W. Leng, "Microstructure and dielectric properties of BaZrxTi1–xO3 ceramics", J. Mater. Sci.: Mater. Electron. **19**, 379-382 (2008); https://doi.org/10.1007/s10854-007-9348-8

[38] R. Farhi, M. El Marssi, A. Simon, and J. Ravez, "A Raman and dielectric study of ferroelectric Ba(Ti1-xZrx)O3 ceramics." Eur. Phys. J. B **9**, 599-604 (1999); https://doi.org/10.1007/s100510050803

[39] T. Maiti, R. Guo, and A. S. Bhalla, "Structure–property phase diagram of BaZrxTi1-xO3 system", J. Amer. Ceram. Soc., **91**(6), 1769-1780 (2008); https://doi.org/10.1111/j.1551-2916.2008.02442.x

[40] B. N. Ezealigo, R. Orrù, C. Elissalde, H. Debéda, U-C. Chung, M. Maglione, and G. Cao, "Influence of the spark plasma sintering temperature on the structure and dielectric properties of BaTi(1-x)ZrxO3 ceramics". Ceramics International **47**, 3614-3625 (2021); https://doi.org/10.1016/j.ceramint.2020.09.210

[41] C. Fu, F. Pan, W. Cai, & X. Deng, "Relaxor behavior of BaZr0.2Ti0.8O3 ceramics with different grains". Integrated Ferroelectrics, **104**, 1-7 (2008); https://doi.org/10.1080/10584580802554844

[42] V. Buscaglia, S. Tripathi, V. Petkov, M. Dapiaggi, M. Deluca, A. Gajović, and Y. Ren, "Average and local atomic-scale structure in BaZrxTi1−xO3 (x = 0.10, 0.20, 0.40) ceramics by high-energy x-ray diffraction and Raman spectroscopy", J. Phys.: Condens. Matter. **26**, 065901 (2014); https://doi.org/10.1088/0953-8984/26/6/065901

[43] B. Zalar, A. Lebar, J. Seliger, R. Blins, V. V. Laguta, and M. Itoh, "NMR studies of disorder in BaTiO3 and SrTiO3," Phys. Rev. B **71**, 064107 (2005); https://doi.org/10.1103/PhysRevB.71.064107

[44] N. Binhayeeniyi, P. Sukwisute, S. Nawae, and N. Muensit, "Energy conversion capacity of barium zirconate titanate", Materials, **13**, 315-324 (2020); https://doi.org/10.3390/ma13020315

[45] S. Singh, K. S. Srikanth and B. Singh, "Pyroelectric performance of [(1-x)Ba0.9Ca0.1TiO3-x(BaSn0.2Ti0.8O3)] lead free ceramics", Ferroelectrics, **526**, 68-75 (2018); https://doi.org/10.1080/00150193.2018.1456277

[46] M. Sharma, S. Singh and B. Singh, "Pyroelectric performance of (Ba0.825+xCa0.175-x)(Ti1-xSnx)O3 lead free ceramics", Ferroelectrics Letters Section, **45**, 76-83 (2018); https://doi.org/10.1080/07315171.2018.1537336

[47] V.V. Laguta, private communications.

[48] O.A. Kovalenko, E.A. Eliseev, Y.O. Zagorodniy, S.D. Škapin, M. M. Kržmanc, L. Demchenko, V.V. Laguta, Z. Kutnjak, D.R. Evans, and A.N. Morozovska. Strain-Gradient and Curvature-Induced

Changes in Domain Morphology of $BaTiO_3$ Nanorods: Experimental and Theoretical Studies, Physical Review Materials **10**, 044409 (2026); https://doi.org/10.1103/b332-gcxc

[49] V. V. Shvartsman, W. Kleemann, J. Dec, Z. K. Xu, S. G. Lu. Diffuse phase transition in BaTi1−xSnxO3 ceramics: An intermediate state between ferroelectric and relaxor behavior, J. Appl. Phys. **99**, 124111 (2006); https://doi.org/10.1063/1.2207828

[50] .A. Santos, J.A. Eiras. Phenomenological description of the diffuse phase transition in ferroelectrics, J Phys: Condens Matter **13**, 11733–11740 (2001); https://doi.org/10.1088/0953-8984/13/50/333

[51] S. L. Bravina, N. V. Morozovsky, A. A. Strokach, "Pyroelectricity: some new research and application aspects", in Material Science and Material Properties for Infrared Optoelectronics, F.F. Sizov (Ed.), Proc. SPIE **3182**, 85-99. (1997). https://doi.org/10.1117/12.280409

[52] S. L. Bravina, N. V. Morozovsky, J. Kulek, B. Hilczer, "Pyroelectric thermowave probing and polarization reversal in TGS/PEO composites". Mol. Cryst. Liq. Cryst., **497**, 109-120 (2008); https://doi.org/10.1080/15421400802458761

[53] A. G. Chynoweth, "Dynamic method for measuring the pyroelectric effect with special reference to barium titanate". J. Appl. Phys., **27**, 78-84 (1956); https://doi.org/10.1063/1.1722201

[54] Y. Kogure, Y. Hiki, "Simultaneous measurement of low-temperature specific heat and thermal conductivity by temperature-wave method". Jpn. J. Appl. Phys., **12**, 814 (1973). https://doi.org/10.1143/JJAP.12.814

[55] M. Oksanen, R. Scholz, L. Fabbri, "Simple thermal wave method for the determination of longitudinal thermal diffusivity of SiC-based fiber". Review of Progress in Quantitative Nondestructive Evaluation. Vol. **17**, Ed. by D.O. Thompson, D.E. Chimenti. (Springer, Boston, 1998); pp.1217-1221 https://doi.org/10.1007/978-1-4615-5339-7_157

[56] A. N. Morozovska, O. S. Pylypchuk, N. V. Morozovsky, E. A. Eliseev, and D. R. Evans. Pyroelectric, electrocaloric and thermoelectric properties of core-shell $Hf_xZr_{1-x}O_2$ nanoparticles: theory and experiment. *ArXiv Preprint* (2026); https://doi.org/10.48550/arXiv.2606.22654

[57] L. E. Cross, Relaxor ferroelectrics. Ferroelectrics **76**, 241 (1987); https://doi.org/10.1080/00150198708016945

[58] E. Courtens, Vogel-Fulcher scaling of the susceptibility in a mixed-crystal proton glass. Physical review letters **52**, 69 (1984); https://doi.org/10.1103/PhysRevLett.52.69

[59] E. Courtens, Scaling dielectric data on $Rb_{1-x}(NH_4)_xH_2PO_4$ structural glasses and their deuterated isomorphs." Phys. Rev. B **33**, 2975 (1986); https://doi.org/10.1103/physrevb.33.2975